\documentclass[a4paper,ngerman,twoside,notitlepage]{article}
\usepackage{amssymb}

\usepackage{float}
\usepackage{makeidx}
\usepackage{amsmath}

\input{tcilatex}
\newcommand{\vekt}[1]{\mbox{\boldmath $#1$\unboldmath}}

\begin{document}

\title{Mean velocity equation\\
for turbulent flow}
\author{J\"{u}rgen Piest}
\maketitle

\begin{abstract}
The hydrodynamic equation derived by Zwanzi-Mori technique of N-particle
statistical mechanics is investigated. This is an attempt to provide
additional information concerning the closure problem of turbulence theory.
The equation is interpreted as mean velocity equation for turbulent fluid
flow. The third-order term of the friction force is calculated. Multilinear
mode coupling theory is applied in order to obtain formulas for the third-
and fourth-order equilibrium time correlation functions appearing in the
expression. The force term is obtained as a convolution integral containing
higher order gradients of the velocity field.
\end{abstract}

\section{Introduction}

\setcounter{equation}{0}

The paper deals with the hydrodynamic equation derived by N-particle
statistical mechanics; especially with the calculation of the third order
term of the friction force.

The motive for the investigation is the closure problem of classical
turbulence theory. It is well known that the problem arises when one
declares, in case of turbulent fluid motion, the hydrodynamic velocity $%
\mathbf{u}$ to be a random variable, and attempts to obtain an equation for
its expectation by just building the expectation over the Navier-Stokes
equation. The non-linear term then gives rise to an additional variable;\
when one formulates an additional equation for this, another variable is
obtained, and so on. Apparently, if the basic equation of motion is
non-linear, this method does not work. Instead, from the point of
probability theory, it would be necessary to find the probability density $f(%
\mathbf{u},t)$\ of the turbulent process, and to calculate the expectation
as:%
\begin{equation}
\langle \mathbf{u}\rangle =\int \mathbf{u\,}f(\mathbf{u},t)d\mathbf{u}
\label{1.1}
\end{equation}

$f$ is functionally dependent on $\mathbf{u}$; thus, (\ref{1.1}) is a
functional integral. As early as 1952, Hopf has formulated this theoretical
approach \cite{Ho}; a short account can be found in \cite{McCo}, ch. 4. - It
became apparent that it is extremely difficult to provide an equation for $%
\langle \mathbf{u}\rangle $\ by elaborating this theory. On the other hand,
the success of direct numerical simulation (see, e. g. \cite{DNS}) leads to
the conclusion that it must be possible to calculate the mean velocitiy on
the basis of the Navier-Stokes equation.

In this situation, it is perhaps reasonable to attempt the derivation of the
mean velocity equation by N-particle statistical mechanics, in order to
provide further information on the nature of the problem. It is obvious that
for investigation of turbulent processes it is not necessary to deal with
the molecules of the fluid. It is the probabilistic structure of statistical
mechanics which constitutes the difference to Navier-Stokes theory.
Statistical mechanics is the basis for describing classical processes
including fluid dynamics; it would be a surprise if turbulent motion would
be an exception. Thus, it is assumed here that it is principally possible to
derive the mean velocity equation by this approach. The complication is that
in a statistical mechanics analysis it is necessary to distinguish between
macroscopic and microscopic parts of the motion, and to handle the latter in
a suitable way. - The Navier-Stokes equation can be derived from statistical
mechanics via the Boltzmann equation by Chapman-Enskog method; see e. g. %
\cite{Hu}. But in the course of the derivation of the Boltzmann equation the
molecular chaos assumption is introduced, which is correct for many
applications including laminar fluid motion, but seems problematic for a
turbulence investigation.

An essential step is defining the mean velocity in the frame of statistical
mechanics. In general, the fluid velocity is defined:%
\begin{equation}
\mathbf{u}=\frac{1}{\rho }\,\langle \mathbf{p}\rangle  \label{1.2}
\end{equation}

$\rho $\ is the mass density, $\mathbf{p}$ the microscopic momentum density.
In this paper, we consider incompressible constant density and temperature
processes. Then $\mathbf{u}$\ is, up to a constant factor, equal to the
expectation of $\mathbf{p}$. Under rather general conditions, the
expectation is equal to the arithmetic mean of a time or space series of the
microscopic quantity. Since for laminar flow fluctuations are microscopic, $%
\mathbf{u}$\ then is the ''point'' velocity of the flow as we usually
understand it. On the other hand, in turbulent flow there appear macroscopic
fluctuations; the arithmetic mean by definition averages over these also;
thus, in this case $\mathbf{u}$\ is already the mean velocity. We have the
peculiar situation that the statistical mechanics $\mathbf{u}$\ is equal the
classical hydrodynamic $\mathbf{u}$\ or \ $\langle \mathbf{u\rangle }$,
depending on whether the flow is laminar or turbulent. Therefore, the
hydrodynamic equation in the form derived by N-particle statistical
mechanics will, by way of trial, be considered the mean velocity equation.

The derivation by means of Zwanzig-Mori projection operator technique of
statistical mechanics (POT) is utilized, in the presentation of Grabert \cite%
{gra}. For the convenience of the reader, some introductory material from an
earlier paper of the author \cite{pi03}\ is repeated. The hydrodynamic
equation obtained by this technique shows a formula for the friction force
containing a local equilibrium time correlation, which is a non-linear
functional of the velocity; presently, there exists no theory from which it
could be calculated. Therefore, it is necessary to develop it into a
functional power series in the velocity;\ this presently restricts the
applicability of the equation to low Reynolds number flow, not very far from
the Navier-Stokes regime. The coefficients of the series now contain total
equlibrium correlation functions, which can be calculated by employing the
multilinear mode coupling theory of Schofield and co-workers \cite{scho},%
\cite{vzs}.

The third-oder term is the first term which furnishes an equation beyond the
Navier-Stokes equation. Time correlation functions of third and fourth oder
appearing in this term are calculated. Finally one obtains an convolution
integral which contains gradients of the velocity up to sixth order. Some
preliminary calculations concerning the design of a test of the theory are
reported.

\bigskip

\section{Hydrodynamic equation}

\setcounter{equation}{0}

This section is a somewhat altered version of section 2 of \cite{pi03}.
There, the description was restricted to stationary processes. In this
paper, I provide the usual time-dependent formulas. The fluid is considered
to be a system of $N$ particles of mass $m$ with positions $\mathbf{y}_{j}$
and velocities $\mathbf{v}_{j}$ which are combined to the phase space matrix 
$z$. Vector components are described by Latin indices, e. g. $\mathbf{y}%
_{j}=\{y_{ja}\}$ . The system is enclosed in a box of Volume $V$ . Later on,
the thermodynamic limit is performed. - A function $g(z)$ is called a phase
space function, or microscopic variable. Especially, we need the space
densities of the conserved quantities mass, energy and momentum $n$, $e$, $%
\mathbf{p}$ which are collected to a 5-element linear matrix $a$. They are
functions of an additional space variable $\mathbf{x}$:%
\begin{equation}
a=\sum_{j=1}^{N}\widetilde{a}_{j}\,\delta (\mathbf{x}-\mathbf{y}_{j})
\label{2.1}
\end{equation}

For the particle functions $\widetilde{a}_{j}$ we have $\widetilde{n}_{j}=m$
, $\widetilde{\mathbf{p}}_{j}=m\mathbf{v}_{j}$ , while the energy function
contains the interparticle potential. $m$ ist the particle mass. Quantities
(like $a$) which are lists of 5 elements are denoted by normal letters while
three-component vectors and tensors are bold. - The quantities $a$ obey the
conservation relations:%
\begin{equation}
\overset{\cdot }{a}=-\nabla \cdot \mathbf{s}  \label{2.2}
\end{equation}

The fluxes $\mathbf{s}$ have the same general structure as the $a$ (\ref{2.1}%
); especially, we have $\mathbf{s}_{1}=\mathbf{p}$ . The time evolution of
any phase space function $A$ is described by the Liouville equation: 
\begin{equation}
\overset{\cdot }{A}\,=\mathcal{L}A  \label{2.2.1}
\end{equation}

$\mathcal{L}$ (often defined as $i\mathcal{L}$)\ is the Liouville operator.
From (\ref{2.2.1}), the formal solution for $A(t)$ given the initial value $%
A $ is:%
\begin{equation}
A(t)=\func{e}^{\mathcal{L}t}A  \label{2.2.2}
\end{equation}

In the statistical model, $z$ and $N$ are considered random variables; that
is, the \ probability density $f(z,N)$ is of grand canonical type. The
ensemble mean value (expectation) of a phase space function $A$ is defined
in the `Heisenberg' picture:%
\begin{equation}
\langle A\rangle (t)=\sum_{N=1}^{\infty }\int dz\,A(z,N,t)f(z,N)  \label{2.3}
\end{equation}

In this formula, $f(z,N)$ is the initial probability distribution, and $%
A(z,N,t)$ is the value of $A$ at time $t$ if the initial positions and
velocities of the particles are described by $z$ . The operation
(integration + Summation) is sometimes indicated by the symbol `tr':%
\begin{equation}
\func{tr}\{\Omega \}=\sum_{N=1}^{\infty }\int dz\Omega (z,N)  \label{2.4}
\end{equation}

Certain probability densities (also called distributions here) are
frequently used in the analysis. One of them is the (total) equilibrium
distribution which corresponds to macroscopic rest: 
\begin{subequations}
\label{2.5}
\begin{gather}
f_{0}=\psi (N)\exp (\Phi _{0}+\beta (\mu N-H(z)))  \label{2.5a} \\
\psi (N)=\frac{1}{N!}(\frac{m}{h})^{3N}  \label{2.5b}
\end{gather}

Here, $h$ is Planck's constant, $\beta =1/(k_{B}T)$ , $k_{B}$ being
Boltzmann's constant and $T$ the temperature, $\mu $ is the chemical
potential which is a function of mass density $\rho =\langle n\rangle $\ and
temperature, and $H(z)$ is Hamilton's function which describes the total
energy of the fluid. For the normalization constant, we have$\ \Phi
_{0}=-\beta PV$, $P$ being the equilibrium pressure. Expectations with
respect to the equilibrium distribution are denoted by $\langle \rangle _{0}$%
. - In case of a simple fluid, the 'relevant probability distribution' of
Grabert's formalism (see \cite{gra}, sec. 2.2) is the local equilibrium
distribution: 
\end{subequations}
\begin{subequations}
\label{2.6}
\begin{gather}
f_{L}(t)=\psi (N)\exp (\Phi (t)-a(z)\circ b(t)),  \label{2.6a} \\
b=\{\beta (\frac{1}{2}u^{2}-\frac{\mu }{m}),\beta ,-\beta \mathbf{u\},}
\label{2.6b} \\
\Phi (t)=-\log (\func{tr}\{\psi \exp (-a\circ b(t))\}).  \label{2.6c}
\end{gather}

Here the symbol $\circ $ is introduced for the operation: Multiplication,
plus Summation over the 5 elements of the linear matrices $a$ , $b$, plus
Integration over geometrical space. The elements of $b$ are called the
conjugate parameters; they are functions of the quantities $\beta $ , $\mu $
and $\mathbf{u}$ which we will sometimes call the thermodynamic parameters,
and which will be considered to be slowly varying functions of space and
time. The $b$\ are defined such that the expectations of the $a$ are
identical to their expectations in local equilibrium: 
\end{subequations}
\begin{equation}
\langle a\rangle =\langle a\rangle _{L}  \label{2.7}
\end{equation}

The POT is a means for separating macroscopic and microscopic parts of the
random variables. It starts by defining the set of phase space functions
which are relevant for the description of the process. For simple fluids,
this set is identified with the densities of conserved variables, $a$\ . A
projection operator \ is defined which projects out of any microscopic
variable $A$ the part which is proportional to the relevant variables. It
reads:

\begin{equation}
\mathcal{P}A=\langle A\rangle _{L}+\langle A\,\delta a\rangle _{L}\circ
\langle \delta a\,\,\delta a\rangle _{L}^{-1}\circ \delta a  \label{2.8}
\end{equation}

Here, $\delta a=a-\langle a\rangle _{L}$; $\langle \rangle _{L}^{-1}$denotes
the inverse of the expectation matrix in the formula. For general
non-stationary flow, the local equilibrium distribution (\ref{2.6a}), and
therefore $\mathcal{P}$, are time-dependent. Then, instead of \cite{pi03}\
(2.11), for the decomposition of $\func{e}^{\mathcal{L}t}$\ the formula \cite%
{gra}\ (2.4.1) is obtained:

\begin{equation}
\func{e}^{\mathcal{L}t}=\func{e}^{\mathcal{L}t}\mathcal{P}(t)\mathcal{+}%
\int_{0}^{t}dt^{\prime }\func{e}^{\mathcal{L}t^{\prime }}\mathcal{P}%
(t^{\prime })(\mathcal{L}-\mathcal{\dot{P}}(t^{\prime }))(1-\mathcal{P}%
(t^{\prime }))\mathcal{G}(t^{\prime },t)+(1-\mathcal{P(}t\mathcal{)})%
\mathcal{G}(0,t)  \label{2.9}
\end{equation}%
\begin{equation}
\mathcal{G}(t^{\prime },t)=\exp _{-}(\int_{t^{\prime }}^{t}dt^{\prime \prime
}\mathcal{L}(1-\mathcal{P}(t^{\prime \prime })))  \label{2.10}
\end{equation}%
$\mathcal{G}(t^{\prime },t)$ \ is a time-ordered exponential Operator which
describes the time dependence of the dissipative part of the equation of
motion. The analysis in \cite{gra} consists in applying (\ref{2.2.2}) to $a$%
, using (\ref{2.9}). By averaging over the initial probability density, and
after some manipulations, Grabert's generalized transport equation \cite{gra}%
, (2.5.17) ist obtained. It is postulated in POT that the initial
probability density is of the form of the 'relevant probability density',
which for simple fluids is defined to be the local equilibrium density (\ref%
{2.6a}). Grabert states that this should not be considered a general
restriction of the method but a means to form the general particle system
into the type specially considered (the simple fluid here); see \cite{gra},
sec. 2.2 . It is shown that in this case the last term in (\ref{2.9})
vanishes after averaging. - Finally the part of the formula pertaining to
the momentum density is taken, and (\ref{1.2}) is used. The result, \cite%
{gra} (8.1.12), (8.1.13), reads with the denotations employed here:

\begin{equation}
\rho (\frac{\partial \mathbf{u}}{\partial t}+\mathbf{u\cdot \nabla
u)=-\nabla }P+\mathbf{\nabla }\cdot \mathbf{R}  \label{2.11}
\end{equation}

\begin{equation}
\mathbf{R}(\mathbf{x,t})=\int_{0}^{t}dt^{\prime }\int d\mathbf{x}^{\prime }%
\mathbf{S}(\mathbf{x},\mathbf{x}^{\prime },t,t^{\prime })\nabla ^{\prime }%
\mathbf{u}(\mathbf{x}^{\prime },t^{\prime })  \label{2.12}
\end{equation}%
\begin{equation}
\mathbf{S}(\mathbf{x},\mathbf{x}^{\prime },t,t^{\prime })=\beta \langle
\lbrack \mathcal{G}(t^{\prime },t)(1-\mathcal{P(}t)\mathbf{s}(\mathbf{x})](1-%
\mathcal{P}(t^{\prime }))\mathbf{s}(\mathbf{x}^{\prime })\rangle
_{L,t^{\prime }}  \label{2.13}
\end{equation}

The microscopic Definition of the pressure $P(x,t)$ \cite{gra}\ (8.4.7) will
not be repeated here. $\langle \,\rangle _{L,t}$\ denotes an expectation
with respect to the local equilibrium distribution at time $t$. Equation (%
\ref{2.11}) has the formal structure of the hydrodynamical equation; but at
the present state of the analysis, it is still an exact equation. It is the
approximations performed later on in the formula for the stress tensor $%
\mathbf{R}$\ which will transform it into an irreversible equation. - In
Addition to (\ref{2.11}), the equation for the particle number density
component of $\langle a\rangle $ \ yields the continuity equation:%
\begin{equation}
\nabla \cdot \mathbf{u=0}  \label{2.14}
\end{equation}

The kernel function $%
\vekt{S}%
$ of the stress Tensor $%
\vekt{R}%
$ is a time correlation function in local equilibrium which\ is a nonlinear
functional of $\mathbf{%
\vekt{u}%
}$. As far as the author knows, presently it is possible to calculate
correlation functions for total equilibrium only. Therefore, it has been
necessary to expand $%
\vekt{S}%
$ into a functional power series in $%
\vekt{u}%
$. As has been explained in section 3 of \cite{pi03}, the expansion can be
done with the set $b(t)$\ of conjugate parameters of the local equilibrium
formula\ (\ref{2.6b}). The expansion is performed at the point $b=b_{0}$,
which corresponds to $\mathbf{u}=0$:

\begin{equation}
b_{0}=\{-\beta \frac{\mu }{m}\,,\,\beta \,,\,0\}  \label{2.15}
\end{equation}

\begin{equation}
b-b_{0}=\{\beta \frac{1}{2}u^{2}\,,\,0\,,\,-\beta \mathbf{u\,\}}
\label{2.16}
\end{equation}%
The power expansion of $S$ reads:

\begin{eqnarray}
\mathbf{S} &=&\mathbf{S}|_{b_{0}}+\frac{\delta \mathbf{S}}{\delta b}%
|_{b_{0}}\ast (b-b_{0})+\frac{1}{2!}\frac{\delta ^{2}\mathbf{S}}{\delta
b\,\delta b}|_{b_{0}}\ast \ast \{(b-b_{0}),(b-b_{0})\}+\cdots  \notag \\
&=&\mathbf{S}^{(0)}+\mathbf{S}^{(1)}+\mathbf{S}^{(2)}+\cdots  \label{2.17}
\end{eqnarray}

The designation of terms in the second row is for later reference. The $\ast 
$\ indicates multiplication, summation over five elements and integration
over space and time. For the present purpose, the expansion is cut after $%
\mathbf{S}^{(2)}$. When these terms are inserted into (\ref{2.12}), one
obtains corresponding parts \ $\mathbf{R}^{(1)}$, \ $\mathbf{R}^{(2)}$, $%
\mathbf{R}^{(3)}$\ \ of $\mathbf{R}$; the upper index again describing the
order in $\mathbf{u}$. The elements of $b_{0}$\ are the conjugate parameters
of the total equilibrium ensemble. Thus, when $S$\ and its derivatives are
taken at $b=b_{0}$, the quantities in the integrand resemble total
equilibrium space-time correlation functions. In \cite{pi03}, these
quantities have been calculated by Kawasaki technique \cite{Ka}; though this
theory still contains certain intuitive elements. In the present state of
the project, for correlation functions the multilinear mode-coupling theory
(MCT) of Schofield and co-workers has been used \cite{scho}, \cite{vzs}. For
the 3-point correlations in $\mathbf{S}^{(1)}$, the results of the two
methodes coincede; for 4-point quantities, there are differences. - In order
to calculate $\mathbf{S}^{(1)}$, it is sufficient to restrict $b-b_{0}$\ to
the last element of (\ref{2.16}), since the first element is of second order
in $\mathbf{u}$; but for $\mathbf{S}^{(2)}$,\ this element of \ $\mathbf{S}%
^{(1)}$ has to be added.

It is an important test for the calculation method that (\ref{2.11}),
correctly approximated, should yield the Navier-Stokes equation. The linear
part $\mathbf{R}^{(1)}$\ of the stress tensor has been calculated by several
authors including Grabert \cite{gra}, sec. 4; the Stokes form of the stress
tensor is obtained, with a microscopic definition of the friction matrix.
There remains another detail of the argumentation: The left-hand side of
equation (\ref{2.11}) is of second order in the velocity. Therefore, in
order for the equation accurately derived, it must be shown that the
second-order part $\mathbf{R}^{(2)}$\ of the stress tensor vanishes. The
present author published two papers in ArXiv \cite{pi07} dealing with this
topic, and finally could show that $\mathbf{R}^{(2)}=0$, so that the
Navier-Stokes equation derives correctly.

\bigskip

\section{Third-order term of the stress tensor}

\setcounter{equation}{0}

Derivation of the third-order term provides considerable effort;\ a computer
algebra system (Mathematica) has been used for most of the calculations. In
order to describe the third-order term, we switch from the stress tensor $%
\mathbf{R}$\ to the friction force $\mathbf{D}=\nabla \cdot \mathbf{R}$. By
introducing the corresponding term of (\ref{2.17}) into (\ref{2.12}), we
obtain a formula for the third-order term $\mathbf{D}_{3}$\ of \ $\mathbf{D}$%
. Again, only the $\mathbf{u}$-part of (\ref{2.16}) has to be considered.
For the detailed investigation, we need the indexed form of the formula:

\begin{eqnarray}
(D_{3})_{a}(\mathbf{x},t) &=&\frac{\beta ^{2}}{2}\,\nabla
_{c}\int_{0}^{t}dt^{\prime }\int d\mathbf{x}^{\prime }\int_{0}^{\infty
}dt^{\prime \prime }\int d\mathbf{x}^{\prime \prime }\int_{0}^{\infty
}dt^{\prime \prime \prime }\int d\mathbf{x}^{\prime \prime \prime }\frac{%
\delta ^{2}S_{abcd}(\mathbf{x},\mathbf{x}^{\prime },t,t^{\prime })}{\delta
b_{e}(\mathbf{x}^{\prime \prime },t^{\prime \prime })\delta b_{f}(\mathbf{x}%
^{\prime \prime \prime },t^{\prime \prime \prime })}|_{b_{0}}\times  \notag
\\
&&\text{ \ \ \ \ \ \ \ \ \ \ \ \ \ \ \ \ \ \ \ \ \ \ \ \ \ \ \ \ \ \ \ \ \ \
\ \ \ \ \ \ \ \ \ \ \ \ \ \ }\times \nabla _{d}^{\prime }\mathbf{u}_{b}(%
\mathbf{x}^{\prime },t^{\prime })\mathbf{u}_{e}(\mathbf{x}^{\prime \prime
},t^{\prime \prime })\mathbf{u}_{f}(\mathbf{x}^{\prime \prime \prime
},t^{\prime \prime \prime })  \label{3.0}
\end{eqnarray}

Latin letters denote indices which run over 3 elements; by contrast, greek
indices run from 1 to 5. - In order to continue, the second-order functional
derivative of $\mathbf{S}$\ is to be calculated. By \cite{pi07}\ (second
paper),\ the first-order derivative consists of two parts. The parametric
part is given by (A7), already specialized to $\mathbf{u}=0$; here we need
the general form:

\begin{equation}
\left[ \frac{\delta S_{abcd}}{\delta b_{\epsilon }(\mathbf{x}^{\prime \prime
},t^{\prime \prime })}\right] _{p}=-\delta (t^{\prime \prime }-t^{\prime
})\beta \,\langle \lbrack \mathcal{G}(t^{\prime },t)\mathcal{Q}(t)s_{ac}(%
\mathbf{x})]\mathcal{Q}(t^{\prime })\delta a_{\epsilon }(\mathbf{x}^{\prime
\prime },t^{\prime })\mathcal{Q}(t^{\prime })s_{bd}(\mathbf{x}^{\prime
})\rangle _{L,t^{\prime }}  \label{3.1}
\end{equation}

\begin{equation}
\mathcal{Q}(t)=1-\mathcal{P}(t)  \label{3.2}
\end{equation}

\begin{equation}
\delta a_{\epsilon }(\mathbf{x},t)=a_{\epsilon }(\mathbf{x})-\langle
a_{\epsilon }(\mathbf{x})\rangle _{L,t}  \label{3.3}
\end{equation}

The functional part is \cite{pi07}\ (second paper) (A14):

\begin{equation}
\left[ \frac{\delta S_{abcd}}{\delta b_{\epsilon }(\mathbf{x}^{\prime \prime
},t^{\prime \prime })}\right] _{f}=\beta \Theta (t^{\prime \prime
}-t^{\prime })\Theta (t-t^{\prime \prime })\langle \lbrack \mathcal{G}%
(t^{\prime },t^{\prime \prime })\mathcal{LP}(t^{\prime \prime }\mathcal{)}%
\delta a_{\epsilon }(\mathbf{x}^{\prime \prime },t^{\prime \prime })\mathcal{%
Q}(t^{\prime \prime }\mathcal{)G}(t^{\prime \prime },t)\mathcal{Q}(t\mathcal{%
)}s_{ac}(\mathbf{x})]\mathcal{Q}(t^{\prime }\mathcal{)}s_{bd}(\mathbf{x}%
^{\prime })\rangle _{L,t^{\prime }}  \label{3.4}
\end{equation}

For later reference, the first-order derivatives are formulated with respect
to all five conjugated parameters (the greek index $\epsilon $\ is used). -
These formulas, as well as the starting formula (\ref{2.13}), contain
operator chains which are, as usual, written like products. The question
arises whether I can, when differentiating a chain, use the product rule; in
other words, if $\mathcal{A}$ and $\mathcal{B}$ are operators, whether it is
correct to write:%
\begin{equation}
\frac{\delta \mathcal{AB}}{\delta b}=\frac{\delta \mathcal{A}}{\delta b}%
\mathcal{B}+\mathcal{A}\frac{\delta \mathcal{B}}{\delta b}  \label{3.5}
\end{equation}

Since I could not find a general proof in the literature, I checked (\ref%
{3.5}) for any two consecutive operators appearing in (\ref{3.1}), (\ref{3.4}%
), and found it correct in all cases. The calculations are partly long and
tedious and will not be shown here. - From (\ref{3.1}), (\ref{3.4}), the
second-order functional derivatives are to be calculated. The formulas
depend on $b$\ by 6 and 8 'factors', respectively; including, in each
formula, the dependence of the local equilibrium probability density. The
corresponding terms are given in Appendix A, (\ref{a.2})\ to \ (\ref{a.17}).
In the course of the calculation, for $\mathcal{G}(t^{\prime },t)$ an
identity formula is needed which can be checked by discretizing it:%
\begin{equation}
\mathcal{Q}(t^{\prime })\mathcal{G}(t^{\prime },t)=\mathcal{Q}(t^{\prime })%
\mathcal{G}(t^{\prime },t)\mathcal{Q}(t)  \label{3.6}
\end{equation}

Three of the terms are found to vanish; thus, the second-order derivative of 
$S$, and therefore the third-order term of $R$ consist of 11 terms. Finally
to be added are parts of the first-order derivative (\ref{3.1}), (\ref{3.4})
which stem from $\epsilon =1$. - For application in (\ref{3.0}), all terms
have to be taken at $b=b_{0}$. This has effects listed below:

- expectations in local equilibrium change into those in total equilibrium,
denoted by $\langle \,\rangle _{0}$;

- projection operators $\mathcal{P}(t)$, $\mathcal{Q}(t)$\ change to
time-independent operators $\mathcal{P}$, $\mathcal{Q}$;

- $\delta a_{e}(x,t)$\ changes to $a_{e}(x)$;

- $\mathcal{G}(t^{\prime },t^{\prime \prime })$\ changes to the exponential
operator $\tilde{g}(t)$\ with $t=t^{\prime \prime }-t^{\prime }$:%
\begin{equation}
\tilde{g}(t)=\func{e}^{\mathcal{LQ}t}  \label{3.7}
\end{equation}

This operator has to be substituted by the similar operator $g(t)$:

\begin{equation}
g(t)=\func{e}^{\mathcal{QL}t}  \label{3.8}
\end{equation}

The reason is that for $g(t)$\ there is an identity which substitutes it
further to the non-projected exponential operator $f(t)=\func{e}^{\mathcal{L}%
t}$. - The substitution succeeds for all 11 terms with the aid of one of the
following identities: 
\begin{subequations}
\label{3.9}
\begin{eqnarray}
\mathcal{Q}\,\tilde{g}(t) &=&g(t)\,\mathcal{Q}  \label{3.9a} \\
\,\tilde{g}(t)\,\mathcal{L} &\mathcal{=}&\mathcal{L\,}g(t)  \label{3.9b} \\
\,\tilde{g}(t)\mathcal{Q}\,\mathcal{L} &\mathcal{=}&-\mathcal{PL+L}g(t)
\label{3.9c}
\end{eqnarray}

Next, we change from the microscopic variables $a$, $\mathbf{s}$\ to the
corresponding ortho-normalized quantities $h$, $\mathbf{r}$; this extracts
from each term the factor $(\frac{\rho }{\beta })^{2}$. Together with the
factor $\beta $\ which appears explicitly in each factor, the pre-factor in (%
\ref{3.0}) changes to $\frac{1}{2}\rho ^{2}\beta $. The gradient operators
appearing in (\ref{3.0}) are transferred to the kernel function; by partial
integration, this alters the sign of the expression. In the formula, we then
have the expressions \ $\nabla _{c}r_{ac}(\mathbf{x})$, $\nabla _{d}^{\prime
}r_{bd}(\mathbf{x}^{\prime })$\ which we denote\ $r_{a}(\mathbf{x})$, $r_{b}(%
\mathbf{x}^{\prime })$ respectively; the distinction to the original $r$'s
is by the number of indices appended. Finally, we change to Fourier space.
The expression now reads:

\end{subequations}
\begin{equation}
(D_{3})_{0}(t)=-\,\frac{1}{2}\rho ^{2}\beta \int_{0}^{t}dt^{\prime
}\int_{0}^{\infty }dt^{\prime \prime }\int_{0}^{\infty }dt^{\prime \prime
\prime }K(r_{0},r_{1}^{\ast },h_{2}^{\ast },h_{3}^{\ast },t,t^{\prime
},t^{\prime \prime },t^{\prime \prime \prime })\mathbf{u}_{1}(t^{\prime })%
\mathbf{u}_{2}(t^{\prime \prime })\mathbf{u}_{3}(t^{\prime \prime \prime })
\label{3.10}
\end{equation}

Here we have introduced number indices. A number index is a combined index
which contains a component index and a wave number variable. Moreover,
double indices contain a summation over the indexed elements and an
integration over the corresponding wave number space, together with the
factor $\frac{1}{(2\pi )^{3}}$. The kernel function $K$\ consists of the 11
Terms presented in Appendix A, (\ref{a.32}) to (\ref{a.42}). For shortness,
I used the abbreviation $\mathbf{\hat{r}=}\mathcal{Q}\mathbf{r}$.

\section{Preparation of the kernel function}

\setcounter{equation}{0}

Starting with this section all expectations are with respect to total
equilibrium. We can therefore omit the $_{0}$ at the expectation symbol. -
We begin by mentioning a well known property of equilibrium correlation
functions. The general form of such a function is:%
\begin{equation}
\langle B_{1}^{(1)}B_{2}^{(2)}\cdots B_{n}^{(n)}\rangle  \label{4.0.1}
\end{equation}

The upper indices just number the microscopic functions $B^{(i)}$, while the
lower ones are the number indices defined in the preceding section. Since
this is a theory with unlimited geometric space, in wave number space the
quantity contains a factor:%
\begin{equation}
(2\pi )^{3}\delta (\mathbf{k}_{1}+\mathbf{k}_{2}+\cdots +\mathbf{k}_{n})
\label{4.0.2}
\end{equation}

This fact simplifies some formulas considerably and will be used several
times in the next sections. - The kernel function of (\ref{3.10}) has to be
brought into a form so that it contains time correlation functions which can
be evaluated by MCT. These are of the general form $\langle \lbrack
f(t)A]B^{\ast }\rangle _{0}$ mit $f(t)=\func{e}^{\mathcal{L}t}$. As a first
step, the projection operators $\mathcal{P}$ and $\mathcal{Q}$\ are to be
substituted. As can be seen in the formulas in the appendix, the terms
contain 0 to 4 of these operators (the operator contained in $\mathbf{\hat{r}%
}$\ remains unaltered). With the reformulations done in the last section,
the operators now read:

\begin{equation}
\mathcal{P\,}A=\langle A\rangle +\langle A\,h_{4}^{\ast }\rangle h_{4}
\label{4.1}
\end{equation}%
\begin{equation}
\mathcal{Q}\,A=A-\langle A\rangle -\langle Ah_{4}^{\ast }\rangle h_{4}
\label{4.2}
\end{equation}

Generally, the component indices contained in the number indices in these
formulas must be greek (running over the full list of 5 elements); it then
would follow that I have to insert $\delta h_{4}$\ instead of \ $h_{4}$. But
as will be seen later, finally all indices reduce to be ''latin'', i. e.
will run over the 3 momentum components. Then, $\langle \,h_{4}\rangle =0$\
, so that formulas (\ref{4.1}), (\ref{4.2}) are correct. - The formulas are
applied in several consecutive steps. After each step, certain auxiliary
formulas are applied in order to simplify the expressions:

\begin{equation}
\begin{array}{c}
\langle h_{1}\hat{r}_{2}\rangle =\langle (\mathcal{Q}h_{1})r_{2}\rangle =0
\\ 
\langle \mathcal{L}A\rangle =0 \\ 
\langle \mathcal{L}g(t)h\rangle =0 \\ 
\langle r\rangle =-\langle \mathcal{L}h\rangle =0 \\ 
\mathcal{L}h_{1}=-\hat{r}_{1}-\langle r_{1}h_{4}\rangle h_{4}%
\end{array}
\label{4.3}
\end{equation}

In this set, the third and fourth formula are special cases of the second. -
The result for the 11 terms, now denoted $K_{1},\cdots ,K_{11}$, is
presented in the appendix, (\ref{a.43}). The $\delta $- and $\theta $%
-factors are temporarily omitted; when the time integrations start they will
be added again.

It is seen that the time correlations in these formulas belong to one of two
groups described by: 
\begin{subequations}
\label{4.4}
\begin{eqnarray}
&&\langle \lbrack g(t)\hat{r}_{0}]B^{\ast }\rangle  \label{4.4a} \\
&&\langle \lbrack \mathcal{L}g(t)h_{0}]B^{\ast }\rangle  \label{4.4b}
\end{eqnarray}

$B$ is a certain microscopic function. The formulas which are still built
with the operator $g(t)$\ are to be transformed into those with $f(t)$. This
will be done in a number of steps. We start with (\ref{4.4a}). First, we
have:

\end{subequations}
\begin{equation}
\langle \lbrack g(t)\hat{r}_{0}]h_{1}^{\ast }\rangle =0  \label{4.5}
\end{equation}

(\ref{3.9a}) and $\mathcal{Q}h_{1}=0$\ have been used. Next, we introduce a
symbol for the memory function of the process:

\begin{equation}
\Gamma _{01}(t)\doteqdot \langle \lbrack g(t)\hat{r}_{0}]\hat{r}_{1}^{\ast
}\rangle  \label{4.6}
\end{equation}

The memory function is, as usual, ''localized'' in time, i. e. approximated
by:%
\begin{equation}
\Gamma _{01}(t)=\delta (t)(2\pi )^{3}\delta (\mathbf{k}_{0}-\mathbf{k}%
_{1})\gamma _{01}  \label{4.7}
\end{equation}

$\gamma _{01}$, when written with component indices $a$, $b$, for
incompressible fluid reduces to:\ 
\begin{equation}
\gamma _{ab}=\nu k_{0}^{2}\delta _{ab}  \label{4.9}
\end{equation}

$\nu $ is the kinematic viscosity, and $\delta _{ab}$ is the Kronecker
symbol. - Next, we need an operator identity described in \cite{zu},
Appendix B2; in the time-dependent form, and applied to the operators
defined in this paper, it reads:

\begin{equation}
g(t)=f(t)-\int_{0}^{t}dt^{\prime }f(t^{\prime })\mathcal{PL}g(t-t^{\prime })
\label{4.10}
\end{equation}

We obtain the correlation function identity, for any microscopic function $B$%
:

\begin{equation}
\langle \lbrack g(t)\hat{r}_{0}]B^{\ast }\rangle =\langle \lbrack f(t)\hat{r}%
_{0}]B^{\ast }\rangle -\int_{0}^{t}dt^{\prime }\Gamma _{04}(t-t^{\prime
})\langle \lbrack f(t^{\prime })h_{4}]B^{\ast }\rangle  \label{4.11}
\end{equation}

After localization of the memory function, we get:

\begin{equation}
\langle \lbrack g(t)\hat{r}_{0}]B^{\ast }\rangle =\langle \lbrack f(t)\hat{r}%
_{0}]B^{\ast }\rangle -\gamma _{04}\langle \lbrack f(t)h_{4}]B^{\ast }\rangle
\label{4.12}
\end{equation}

The first term on the rhs of (\ref{4.11}) is still to be transformed. The
fifth formula (\ref{4.3}) together with (\ref{2.2.2}) and $\omega
_{04}=\langle r_{0}h_{4}^{\ast }\rangle $\ yields:%
\begin{equation}
\hat{r}_{0}=-\partial _{t}h_{0}-\omega _{04}h_{4}  \label{4.14}
\end{equation}

We obtain:

\begin{equation}
\langle \lbrack g(t)\hat{r}_{0}]B^{\ast }\rangle =-\partial _{t}\langle
\lbrack f(t)h_{0}]B^{\ast }\rangle -\kappa _{04}\langle \lbrack
f(t)h_{4}]B^{\ast }\rangle  \label{4.15}
\end{equation}%
\begin{equation}
\kappa _{04}=\omega _{04}+\gamma _{04}  \label{4.16}
\end{equation}

We denote with $C_{n}$ die time correlation functions which can be
calculated by MCT:%
\begin{equation}
(C_{n})_{01\cdots n-1}(t)\doteqdot \langle \lbrack f(t)h_{0}]h_{1}^{\ast
}\cdots h_{n-1}^{\ast }\rangle  \label{4.17}
\end{equation}

Then, we have the following applications of (\ref{4.15}):

\begin{equation}
\langle \lbrack g(t)\hat{r}_{0}]h_{1}^{\ast }h_{2}^{\ast }\rangle =-\partial
_{t}(C_{3})_{012}(t)-\kappa _{08}(C_{3})_{812}(t)  \label{4.18}
\end{equation}

\begin{equation}
\langle \lbrack g(t)\hat{r}_{0}]h_{1}^{\ast }h_{2}^{\ast }h_{3}^{\ast
}\rangle =-\partial _{t}(C_{4})_{0123}(t)-\kappa _{08}(C_{4})_{8123}(t)
\label{4.19}
\end{equation}

Somewhat more complicated is the transformation if $B$\ contains $\hat{r}$\
as a factor. Again, (\ref{4.14}) is applied, together with $\omega
_{04}^{\ast }=-\omega _{04}$:

\begin{align}
& \langle \lbrack g(t)\hat{r}_{0}]\hat{r}_{1}^{\ast }h_{2}^{\ast }\rangle
=-\partial _{t}\langle \lbrack f(t)h_{0}]\hat{r}_{1}^{\ast }h_{2}^{\ast
}\rangle -\kappa _{04}\langle \lbrack f(t)h_{4}]\hat{r}_{1}^{\ast
}h_{2}^{\ast }\rangle  \notag \\
& =\partial _{t}\langle \lbrack f(t)h_{0}](\partial _{t}h_{1}^{\ast
})h_{2}^{\ast }\rangle -\omega _{17}\partial _{t}\langle \lbrack
f(t)h_{0}]h_{7}^{\ast }h_{2}^{\ast }\rangle +\kappa _{04}(\langle \lbrack
f(t)h_{4}](\partial _{t}h_{1}^{\ast })h_{2}^{\ast }\rangle -\omega
_{17}\langle \lbrack f(t)h_{4}]h_{7}^{\ast }h_{2}^{\ast }\rangle )
\label{4.20}
\end{align}

The first and the third term need further transformation. Because of the
symmetries of (\ref{3.10}), the following formula is correct if it is used
only within this expression:

\begin{equation}
(\partial _{t}h_{1}^{\ast })h_{2}^{\ast }=\tfrac{1}{2}((\partial
_{t}h_{1}^{\ast })h_{2}^{\ast }+h_{1}^{\ast }\partial _{t}h_{2}^{\ast })=%
\tfrac{1}{2}\partial _{t}(h_{1}^{\ast }h_{2}^{\ast })  \label{4.21}
\end{equation}

We obtain, for the correlation function (\ref{4.20}):

\begin{align}
& \langle \lbrack g(t)\hat{r}_{0}]\hat{r}_{1}^{\ast }h_{2}^{\ast }\rangle = 
\notag \\
& =-\tfrac{1}{2}\partial _{tt}(C_{3})_{012}(t)-\omega _{17}\partial
_{t}(C_{3})_{072}(t)-\kappa _{04}(\tfrac{1}{2}\partial
_{t}(C_{3})_{412}(t)+\omega _{17}(C_{3})_{472}(t))  \label{4.22}
\end{align}

We now switch to time correlations belonging to group (\ref{4.4b}). We need
an operator identity similar to (\ref{4.10}) which again can be found in %
\cite{zu}, Appendix 2B. Applied to this paper, it reads:

\begin{equation}
g(t)=f(t)-\int_{0}^{t}dt^{\prime }g(t^{\prime })\mathcal{PL}f(t-t^{\prime })
\label{4.23}
\end{equation}

We obtain a formula similar to (\ref{4.11}):

\begin{equation}
\langle \lbrack \mathcal{L}g(t)h_{0}]B^{\ast }\rangle =\langle \lbrack 
\mathcal{L}f(t)h_{0}]B^{\ast }\rangle -\int_{0}^{t}dt^{\prime }\langle
\lbrack \mathcal{L}f(t-t^{\prime })h_{0}]h_{4}^{\ast }\rangle \langle
\lbrack \mathcal{L}g(t^{\prime })h_{4}]B^{\ast }\rangle  \label{4.24}
\end{equation}

For shortness, we introduce some new denotations:

\begin{equation}
\Psi _{0}(t)=\langle \lbrack \mathcal{L}g(t)h_{0}]B^{\ast }\rangle
\label{4.25}
\end{equation}%
\begin{equation}
\Phi _{0}(t)=\langle \lbrack \mathcal{L}f(t)h_{0}]B^{\ast }\rangle =\partial
_{t}\langle \lbrack f(t)h_{0}]B^{\ast }\rangle  \label{4.26}
\end{equation}

Then, applying (\ref{4.14}), (\ref{4.24}) reads:

\begin{equation}
\Psi _{0}(t)=\Phi _{0}(t)-\int_{0}^{t}dt^{\prime }\partial
_{t}(C_{2})_{04}(t-t^{\prime })\Psi _{4}(t^{\prime })  \label{4.27}
\end{equation}

This is an integral equation for $\Psi $\ which can be formally solved by
temporarily switching to Laplace space. The result is:

\begin{equation}
\Psi _{0}(t)=\Phi _{0}(t)+\kappa _{04}\int_{0}^{t}dt^{\prime }\Phi
_{4}(t^{\prime })  \label{4.28}
\end{equation}

Taking (\ref{4.26}) into account, we can integrate (\ref{4.24}):

\begin{equation}
\langle \lbrack \mathcal{L}g(t)h_{0}]B^{\ast }\rangle =\partial _{t}\langle
\lbrack f(t)h_{0}]B^{\ast }\rangle +\kappa _{04}(\langle \lbrack
f(t)h_{4}]B^{\ast }\rangle -\langle h_{4}B^{\ast }\rangle )  \label{4.29}
\end{equation}

This is the basic formula for time correlations of form (\ref{4.4b}). Taking 
$B=h_{1}$, we have:

\begin{equation}
\langle \lbrack \mathcal{L}g(t)h_{0}]h_{1}^{\ast }\rangle =\partial
_{t}(C_{2})_{01}(t)+\kappa _{07}((C_{2})_{71}(t)-\delta _{71})  \label{4.30}
\end{equation}

The lowest-order non-projected time correlation function $C_{2}$\ obeys a
simple differential equation (\cite{pi07}, second paper, (4.6)):

\begin{equation}
\partial _{t}(C_{2})_{01}(t)=-\kappa _{07}(C_{2})_{71}(t)  \label{4.31}
\end{equation}

This is introduced into (\ref{4.30}):

\begin{equation}
\langle \lbrack \mathcal{L}g(t)h_{0}]h_{1}^{\ast }\rangle =-\kappa _{01}
\label{4.32}
\end{equation}

The lowest-order time correlation of type (\ref{4.4b}) is a constant. - For
the next-higher order, we find from (\ref{4.29}):

\begin{equation}
\langle \lbrack \mathcal{L}g(t)h_{0}]h_{1}^{\ast }h_{2}^{\ast }\rangle
=\partial _{t}(C_{3})_{012}(t)+\kappa _{07}((C_{3})_{712}(t)-j_{712})
\label{4.33}
\end{equation}%
\begin{equation}
(j_{3})_{012}=(C_{3})_{012}(t=0)=\langle h_{0}h_{1}^{\ast }h_{2}^{\ast
}\rangle  \label{4.34}
\end{equation}

Again the situation is somewhat more complicated if $B$\ contains the
projected flux $\hat{r}$. For the lowest order, we obtain, by applying (\ref%
{4.29}):%
\begin{equation}
\langle \lbrack \mathcal{L}g(t)h_{0}]\hat{r}_{1}^{\ast }h_{2}^{\ast }\rangle
=\partial _{t}\langle \lbrack f(t)h_{0}]\hat{r}_{1}^{\ast }h_{2}^{\ast
}\rangle +\kappa _{04}(\langle \lbrack f(t)h_{4}]\hat{r}_{1}^{\ast
}h_{2}^{\ast }\rangle -\langle h_{4}\hat{r}_{1}^{\ast }h_{2}^{\ast }\rangle )
\label{4.35}
\end{equation}

We apply (\ref{4.14}):

\begin{eqnarray}
\langle \lbrack \mathcal{L}g(t)h_{0}]\hat{r}_{1}^{\ast }h_{2}^{\ast }\rangle
&=&-\partial _{t}\langle \lbrack f(t)h_{0}](\partial _{t}h_{1}^{\ast
})h_{2}^{\ast }\rangle +\omega _{17}\partial _{t}\langle \lbrack
f(t)h_{0}]h_{7}^{\ast }h_{2}^{\ast }\rangle +  \notag \\
&&+\kappa _{04}(-\langle \lbrack f(t)h_{4}](\partial _{t}h_{1}^{\ast
})h_{2}^{\ast }\rangle +\omega _{17}\langle \lbrack f(t)h_{4}]h_{7}^{\ast
}h_{2}^{\ast }\rangle -\langle h_{4}\hat{r}_{1}^{\ast }h_{2}^{\ast }\rangle )
\label{4.36}
\end{eqnarray}

We introduce a notation for the static correlation appearing in the formula:

\begin{equation}
(s_{3})_{124}=\langle \hat{r}_{1}h_{2}h_{4}^{\ast }\rangle =-\langle \hat{r}%
_{1}h_{2}h_{4}^{\ast }\rangle ^{\ast }=-\langle h_{4}\hat{r}_{1}^{\ast
}h_{2}^{\ast }\rangle  \label{4.36a}
\end{equation}

Finally, we can apply a symmetry consideration as in (\ref{4.20}); by using (%
\ref{4.21}) we obtain:

\begin{eqnarray}
\langle \lbrack \mathcal{L}g(t)h_{0}]\hat{r}_{1}^{\ast }h_{2}^{\ast }\rangle
&=&\tfrac{1}{2}\partial _{tt}(C_{3})_{012}(t)+\omega _{17}\partial
_{t}(C_{3})_{072}(t)+  \notag \\
&&+\kappa _{04}(\tfrac{1}{2}\partial _{t}(C_{3})_{412}(t)+\omega
_{17}(C_{3})_{472}(t)+(s_{3})_{124})  \label{4.37}
\end{eqnarray}

For the next-higher order correlation we have, by (\ref{4.29}):

\begin{equation}
\langle \lbrack \mathcal{L}g(t)h_{0}]\hat{r}_{1}^{\ast }h_{2}^{\ast
}h_{3}^{\ast }\rangle =\partial _{t}\langle \lbrack f(t)h_{0}]\hat{r}%
_{1}^{\ast }h_{2}^{\ast }h_{3}^{\ast }\rangle +\kappa _{04}(\langle \lbrack
f(t)h_{4}]\hat{r}_{1}^{\ast }h_{2}^{\ast }h_{3}^{\ast }\rangle -\langle h_{4}%
\hat{r}_{1}^{\ast }h_{2}^{\ast }h_{3}^{\ast }\rangle )  \label{4.38}
\end{equation}

Introduction of (\ref{4.14}) yields:

\begin{eqnarray}
\langle \lbrack \mathcal{L}g(t)h_{0}]\hat{r}_{1}^{\ast }h_{2}^{\ast
}h_{3}^{\ast }\rangle &=&-\partial _{t}\langle \lbrack f(t)h_{0}](\partial
_{t}h_{1}^{\ast })h_{2}^{\ast }h_{3}^{\ast }\rangle +\omega _{17}\partial
_{t}\langle \lbrack f(t)h_{0}]h_{7}^{\ast }h_{2}^{\ast }h_{3}^{\ast }\rangle
)  \notag \\
&&+\kappa _{04}(-\langle \lbrack f(t)h_{4}](\partial _{t}h_{1}^{\ast
})h_{2}^{\ast }h_{3}^{\ast }\rangle +\omega _{17}\langle \lbrack
f(t)h_{4}]h_{7}^{\ast }h_{2}^{\ast }h_{3}^{\ast }\rangle -\langle h_{4}\hat{r%
}_{1}^{\ast }h_{2}^{\ast }h_{3}^{\ast }\rangle )  \label{4.39}
\end{eqnarray}

The symmetry consideration which is correct if it is applied within the
integral formula (\ref{3.10}) gives:

\begin{equation}
(\partial _{t}h_{1}^{\ast })h_{2}^{\ast }h_{3}^{\ast }=\tfrac{1}{3}%
((\partial _{t}h_{1}^{\ast })h_{2}^{\ast }h_{3}^{\ast }+h_{1}^{\ast
}(\partial _{t}h_{2}^{\ast })h_{3}^{\ast }+h_{1}^{\ast }h_{2}^{\ast
}\partial _{t}h_{3}^{\ast })=\tfrac{1}{3}\partial _{t}(h_{1}^{\ast
}h_{2}^{\ast }h_{3}^{\ast })  \label{4.40}
\end{equation}

Again a denotation for the static correlation is introduced:

\begin{equation}
(s_{4})_{1234}=\langle \hat{r}_{1}h_{2}h_{3}h_{4}^{\ast }\rangle =-\langle 
\hat{r}_{1}h_{2}h_{3}h_{4}^{\ast }\rangle ^{\ast }=-\langle h_{4}\hat{r}%
_{1}^{\ast }h_{2}^{\ast }h_{3}^{\ast }\rangle  \label{4.41}
\end{equation}

The final form for (\ref{4.38}) is:

\begin{eqnarray}
\langle \lbrack \mathcal{L}g(t)h_{0}]\hat{r}_{1}^{\ast }h_{2}^{\ast
}h_{3}^{\ast }\rangle &=&\tfrac{1}{3}\partial _{tt}(C_{4})_{0123}(t)+\omega
_{17}\partial _{t}(C_{4})_{0723}(t)+  \notag \\
&&+\kappa _{04}(\tfrac{1}{3}\partial _{t}(C_{4})_{4123}(t)+\omega
_{17}(C_{4})_{4723}(t)+(s_{4})_{1234})  \label{4.42}
\end{eqnarray}

The formulas obtained in this section are introduced into the terms (\ref%
{a.43}). The Dirac and Heaviside functions omitted are added again. The
kernel function $K$\ in (\ref{3.10})\ now consists of 5 main parts:%
\begin{eqnarray}
K &=&\delta (t^{\prime \prime \prime }-t^{\prime })\delta (t^{\prime \prime
}-t^{\prime })M_{1}  \notag \\
&&+\delta (t^{\prime \prime }-t^{\prime })\Theta (t^{\prime \prime \prime
}-t^{\prime })\Theta (t-t^{\prime \prime \prime })M_{2}  \notag \\
&&+\delta (t^{\prime \prime \prime }-t^{\prime \prime })\Theta (t^{\prime
\prime }-t^{\prime })\Theta (t-t^{\prime \prime })\delta (t^{\prime \prime
}-t^{\prime })M_{3}  \label{4.43} \\
&&+\Theta (t^{\prime \prime \prime }-t^{\prime \prime })\Theta (t-t^{\prime
\prime \prime })\Theta (t^{\prime \prime }-t^{\prime })\Theta (t-t^{\prime
\prime })\delta (t^{\prime \prime }-t^{\prime })M_{4}  \notag \\
&&+\Theta (t^{\prime \prime \prime }-t^{\prime })\Theta (t^{\prime \prime
}-t^{\prime \prime \prime })\Theta (t^{\prime \prime }-t^{\prime })\Theta
(t-t^{\prime \prime })\delta (t^{\prime \prime \prime }-t^{\prime })M_{5} 
\notag
\end{eqnarray}

The coefficients $M_{1},\cdots ,M_{5}$\ are presented in the appendix, (\ref%
{a.44}). We denote:%
\begin{equation}
(j_{2})_{12}=\langle h_{1}^{\ast }h_{2}^{\ast }\rangle  \label{4.43a}
\end{equation}

When (\ref{a.44}) is inserted into (\ref{3.10}), some of the time integrals
can be executed. Moreover, we can apply an approximation since the time
scale of the $\mathbf{u}$-factors is much larger than that of the
equilibrium time correlations. The upper limits of integrals over
correlation functions can then be extended to infinity. The formula now
reads:

\begin{eqnarray}
(D_{3})_{0}(t) &=&-\,\frac{1}{2}\rho ^{2}\beta \mathbf{u}_{1}(t)\mathbf{u}%
_{2}(t)\mathbf{u}_{3}(t)\times  \notag \\
&&\times (\int_{0}^{\infty }dt^{\prime }\Phi 1_{0123}(t^{\prime
})+\lim_{t\rightarrow \infty }\int_{0}^{t}dt^{\prime }\int_{0}^{t-t^{\prime
}}dt^{\prime \prime }\Phi 2_{0123}(t^{\prime },t^{\prime \prime }))
\label{4.44}
\end{eqnarray}

with $\Phi 1=M_{1}+M_{3}$\ and $\Phi 2=M_{2}+M_{4}+M_{5}$. The detailed
expressions are given in the appendix, (\ref{a.45}). - This is the part of
the third-order term which stems from the term $\mathbf{S}^{(2)}$\ of the
kernel function in expression (\ref{2.12}) for the stress tensor. As has
been mentioned earlyer, still to be added is the part of the second-order
term which derives from (\ref{3.2}), (\ref{3.5}) for $\epsilon =1$.

\bigskip

\section{Relevant part of the second-order term}

\setcounter{equation}{0}

When one treats the second-order term in the same way which led to (\ref%
{3.10}), one obtains:\ 

\begin{equation}
(D_{2})_{0}(t)=-\frac{1}{2}\rho ^{\frac{3}{2}}\beta ^{\frac{1}{2}}\left( 
\frac{\partial p}{\partial \rho }\right) ^{-\frac{1}{2}}\,\int_{0}^{t}dt^{%
\prime }\int_{0}^{\infty }dt^{\prime \prime }(K_{2})_{012}\left( t,t^{\prime
},t^{\prime \prime }\right) u_{1}(t^{\prime })uq(t^{\prime \prime })
\label{5.1}
\end{equation}

In this section, the number index 2\ contains a component index which has
the fixed value 1, if not stated otherwise. The pre-factor contains a part
which results from switching to normalized variables in the correlation
functions. $uq$\ is the Fourier transform of $u^{2}$\ for the wave number
variable $\mathbf{k}^{\prime \prime }$. The kernel function $K_{2}$\ is the
sum of (\ref{3.1}) and (\ref{3.4}), taken at $b=b_{0}$ for $\epsilon =1$; (%
\ref{3.9a}) has been used to substitute the operator $\tilde{g}$:

\begin{align}
(K_{2})_{012}\left( t,t^{\prime },t^{\prime \prime }\right) & =-\delta
(t^{\prime \prime }-t^{\prime })\,\langle \lbrack g(t-t^{\prime })\hat{r}%
_{0}]\hat{r}_{1}^{\ast }\delta h_{2}^{\ast }\rangle +  \notag \\
& \text{ \ \ \ \ }+\Theta (t^{\prime \prime }-t^{\prime })\Theta
(t-t^{\prime \prime })\langle \lbrack g(t^{\prime \prime }-t^{\prime })%
\mathcal{QLP}\delta h_{2}^{\ast }g(t-t^{\prime \prime })\hat{r}_{0}]\hat{r}%
_{1}^{\ast }\rangle  \label{5.2}
\end{align}

For the first term, we have:%
\begin{equation}
\langle \lbrack g(t-t^{\prime })\hat{r}_{0}]\hat{r}_{1}^{\ast }\delta
h_{2}^{\ast }\rangle =\langle \lbrack g(t-t^{\prime })\hat{r}_{0}]\hat{r}%
_{1}^{\ast }h_{2}^{\ast }\rangle -\langle \lbrack g(t-t^{\prime })\hat{r}%
_{0}]\hat{r}_{1}^{\ast }\rangle \langle h_{2}^{\ast }\rangle  \label{5.3}
\end{equation}

The operator $\mathcal{P}$\ in the second term in (\ref{5.2}) is evaluated:

\begin{align}
\langle \lbrack g& (t^{\prime \prime }-t^{\prime })\mathcal{QLP}\delta
h_{2}^{\ast }g(t-t^{\prime \prime })\hat{r}_{0}]\hat{r}_{1}^{\ast }\rangle 
\notag \\
\text{ \ }& =-\langle \lbrack g(t-t^{\prime \prime })\hat{r}_{0}]h_{2}^{\ast
}h_{3}^{\ast }\rangle \langle \lbrack g(t^{\prime \prime }-t^{\prime })\hat{r%
}_{3}]\hat{r}_{1}^{\ast }\rangle  \label{5.4}
\end{align}

The identity $\mathcal{QL}\delta h_{3}=-\mathcal{Q}r_{3}=-\hat{r}_{3}$\ has
been used. Originally, the full form (\ref{4.1}) with $\delta h$\ instead of 
$h$\ had to be used in the first factor;\ but it is found that the terms
with the constant parts \ $\langle h_{2}\rangle $, $\langle h_{3}\rangle $\
vanish, so that the final form is (\ref{5.4}). - For further preparation the
formulas (\ref{4.6}) and (\ref{4.7}), (\ref{4.18}), (\ref{4.22}) are used,
integration over $t^{\prime \prime }$\ and an approximation because of
different time scales is performed:

\begin{equation}
(D_{2})_{0}(t)=-\frac{1}{2}\rho ^{\frac{3}{2}}\beta ^{\frac{1}{2}}\left(
\partial _{\rho }P|_{\beta }\right) ^{-\frac{1}{2}}\,u_{1}(t)uq(t)\int_{0}^{%
\infty }dt^{\prime }(K_{2})_{012}\left( t^{\prime }\right)  \label{5.5}
\end{equation}

\ $K_{2}$, now a function of $t^{\prime }$\ only, is given in the appendix, (%
\ref{a.46}).

(\ref{5.5}) \ has to be added to (\ref{4.44})\ in order to obtain the full
third-order term $D_{3}$\ of the friction force:

\begin{eqnarray}
(D_{3})_{0}(t) &=&-\,\frac{1}{2}\rho ^{2}\beta (\mathbf{u}_{1}(t)\mathbf{u}%
_{2}(t)\mathbf{u}_{3}(t)\times  \notag \\
&&\times (\int_{0}^{\infty }dt^{\prime }\Phi 1_{0123}(t^{\prime
})+\lim_{t\rightarrow \infty }\int_{0}^{t}dt^{\prime }\int_{0}^{t-t^{\prime
}}dt^{\prime \prime }\Phi 2_{0123}(t^{\prime },t^{\prime \prime }))  \notag
\\
&&+(\rho \beta \partial _{\rho }P|_{\beta })^{-\frac{1}{2}%
}\,u_{1}(t)uq(t)\int_{0}^{\infty }dt^{\prime }(K_{2})_{012}\left( t^{\prime
}\right) )  \label{5.6}
\end{eqnarray}

To complete the derivation, the equilibrium correlation functions $C_{3}$\
and $C_{4}$ appearing in the coefficients have to be calculated. The MCT has
been applied for this purpose \cite{scho},\cite{vzs}. A formula for $C_{3}$\
is explicitly given in \cite{scho}, while $C_{4}$ has been calculated with
the aid of the rules given in these papers. This will be described in next
section.

\section{Multilinear mode coupling theory}

\setcounter{equation}{0}

We refer to the paper by van Zon and Schofield \cite{vzs}. We continue to
use the ortho-normalized microscopic variables $h$ defined in sec. 3. In %
\cite{vzs}, multiple variables $A_{\alpha }$\ are employed which are
products of the single variables:%
\begin{equation}
A_{\alpha }=h_{1}h_{2}\cdots h_{\alpha }  \label{6.1}
\end{equation}

Greek indices, in addtion to describing a component index running over 5
elements, are used in this section in connection with the multiple
variables. Depending on the context, they\ denote either the order of the \
multiple variable or the full set of number indices:%
\begin{equation}
\alpha =\left\{ 1,2,\cdots ,\alpha \right\}  \label{6.2}
\end{equation}

Orthogonal multiple variables $Q_{\alpha }$\ and projection operators $%
\mathcal{P}_{\alpha }$ are successively defined (do not mix up these symbols
with the (non-indexed) $\mathcal{P}$ and $Q$\ used in earlier sections):%
\begin{equation}
\left. 
\begin{array}{c}
Q_{0}=1 \\ 
Q_{\alpha }=(1-\sum_{\nu =0}^{\alpha -1}\mathcal{P}_{\nu })A_{\alpha } \\ 
\mathcal{P}_{0}X=\langle X\rangle \\ 
\mathcal{P}_{\alpha }X=\langle X\,Q_{\alpha }\rangle Q_{\alpha }%
\end{array}%
\right\}  \label{6.3}
\end{equation}

A greek symbol appearing twice includes summation over each of the component
indices and integration over each of the wave numbers;\ for each wave number
including a factor \ $(2\pi )^{-3}$. The last of the formulas (\ref{6.3}) is
somewhat simpler than \cite{vzs}\ (6), owing to the fact that our basic
variables $h$\ are normalized. - An important property of the theory is that
it is formulated before performing the thermodynamic limit; that is, there
is a finite number of particles, $N$. Finally, the limit $N\rightarrow
\infty $\ is taken. In the formulas appear terms which are proportional to
different powers of $N$; in the limit, only those with the highest power of $%
N$\ 'survive'. In this paper, we will provide only the final formulas
obtained after performing the thermodynamic limit. - The object of the
theory are time correlations of the $Q$:%
\begin{equation}
G_{\alpha \beta }=\langle Q_{\alpha }(t)Q_{\beta }^{\ast }\rangle
\label{6.4}
\end{equation}

Again, formula (\ref{6.4}) is somewhat simpler than \cite{vzs} (15). In the
correlations appearing in $D_{3}$\ and $D_{2}$, we have $\alpha =1$; for $%
\beta =1$, (\ref{6.4}) is directly the two-point correlation $C_{2}$, while
the higher-order correlations $C_{3}$\ and $C_{4}$\ are obtained from (\ref%
{6.4}) for $\beta =2$ and $3$, respectively. In the latter cases, formula %
\cite{vzs}\ (26) (given in Laplace space) has to be applied. Since $\beta
\neq \alpha $, the first term vanishes. The case $\beta =2$\ is analyzed in %
\cite{vzs}\ as an example. The $N$-ordering analysis shows that only the
second term of the formula survives. For $\beta =3$, the second and certain
parts of the third term survive. But the detailed analysis shows that the
third term contains a dimensionless combination of thermodynamic parameters
which is $\ll 1$, so that finally both cases are described by the same
formula:%
\begin{equation}
G_{\alpha \beta }(\zeta )=G_{\alpha \alpha ^{\prime }}(\zeta )M_{\alpha
^{\prime }\beta ^{\prime }}(\zeta )G_{\beta ^{\prime }\beta }(\zeta )
\label{6.5}
\end{equation}%
\begin{equation}
M_{\alpha \beta }(\zeta )=\langle \dot{Q}_{\alpha }Q_{\beta }^{\ast }\rangle
-\Gamma _{\alpha \beta }(\zeta )  \label{6.6}
\end{equation}

Certain simplifications (compared with \cite{vzs}\ (26)) stem from the fact
that for the quantities considered $\beta \neq \alpha $. $\zeta $\ is the
independent variable of Laplace space. $\alpha ^{\prime }$\ describes an
index set of the order $\alpha $\ with the same wave numbers as $\alpha $\
but different component indices.

$\Gamma _{\alpha \beta }$\ is the time correlation of the fluctuating forces
defined in \cite{vzs}\ (11). These quantities describe the friction effects
of the fluid. Generally, $\Gamma _{\alpha \beta }(z)$\ is replaced by its
value at $z=0$ ('localization' in time). The lowest order quantity, $\alpha
=\beta =1$, for small wave numbers resembles the Green-Kubo expressions. In %
\cite{vzs}, there is no prescription for the higher order $\Gamma _{\alpha
\beta }$. We refer to the older correlation function theory by Kawasaki \cite%
{k}, where these quantities are neglected. - In addition to (\ref{6.5}), (%
\ref{6.6}), it is shown (\cite{vzs} (25)) that, after transforming back to
the time domain, for any $\alpha $, in highest $N$-order $G_{\alpha \alpha
^{\prime }}(t)$\ factorizes into the product of the corresponding order-one
quantities. - Essentially, (\ref{6.5}), (\ref{6.6}) express the higher order
correlations in terms of $C_{2}$; to calculate $C_{2}$, an additional
formula is used (see below).

For $C_{3}$, the formulas derived in \cite{vzs}\ are (30), (31);\ with the
denotations of the present paper and in terms of number indices, they read:%
\begin{equation}
(C_{3})_{123}(t)=(C_{2})_{14}(t)(j_{3})_{423}+(G_{12})_{123}(t)  \label{6.7}
\end{equation}%
\begin{equation}
(G_{12})_{123}(t)=-\int_{0}^{t}dt^{\prime }\,(C_{2})_{14}(t-t^{\prime
})(s_{3})_{456}(C_{2})_{25}(t^{\prime })(C_{2})_{36}(t^{\prime })
\label{6.8}
\end{equation}%
\begin{equation}
(s_{3})_{123}=\langle \hat{r}_{1}h_{2}^{\ast }h_{3}^{\ast }\rangle
\label{6.10}
\end{equation}

$j_{3}$\ is defined in (\ref{4.34}).\ Flux densities $r$\ and projected flux
densities $\hat{r}$\ are defined in\ (4.3). $s_{3}$\ is, by definition,
different from the quantity defined in (4.36). But after the factor
extraction described at the beginning of the next section, the difference
will vanish. We will therefore use the same notation. - The formulas for $%
C_{4}$ obtained from (\ref{6.5}), (\ref{6.6}) for $\beta =3$\ read, after
some manipulations:%
\begin{equation}
(C_{4})_{1234}(t)=((C_{3})_{156}(t)-(j_{3})_{156})\Phi
_{23456}+(C_{2})_{17}(t)(j_{4})_{7234}+(G_{13})_{1234}(t)  \label{6.12}
\end{equation}

\begin{equation}
(G_{13})_{1234}(t)=-\int_{0}^{t}dt^{\prime }\,(C_{2})_{15}(t-t^{\prime
})(\sigma _{4})_{5678}(C_{2})_{26}(t^{\prime })(C_{2})_{37}(t^{\prime
})(C_{2})_{48}(t^{\prime })  \label{6.13}
\end{equation}%
\begin{equation}
\Phi
_{12345}=(j_{5})_{12345}-(j_{3})_{123}(j_{2})_{45}-(j_{4})_{1237}(j_{3})_{745}
\label{6.14}
\end{equation}%
\begin{equation}
(\sigma _{4})_{1234}=(s_{4})_{1234}-\Phi _{23456}(s_{3})_{156}  \label{6.15}
\end{equation}%
\begin{equation}
\left. 
\begin{array}{c}
(j_{2})_{12}=\langle h_{1}h_{2}\rangle \\ 
(j_{3})_{123}=\langle h_{1}h_{2}h_{3}\rangle \\ 
(j_{4})_{1234}=\langle h_{1}h_{2}^{\ast }h_{3}^{\ast }h_{4}^{\ast }\rangle
\\ 
(j_{5})_{12345}=\langle h_{1}h_{2}h_{3}^{\ast }h_{4}^{\ast }h_{5}^{\ast
}\rangle \\ 
(s_{4})_{1234}=\langle \hat{r}_{1}h_{2}^{\ast }h_{3}^{\ast }h_{4}^{\ast
}\rangle%
\end{array}%
\right\}  \label{6.16}
\end{equation}

The choice of conjugate complex variables in (\ref{6.16}) looks somewhat
arbitrary; the quantities are defined as they appear in (\ref{6.12}) to (\ref%
{6.15}). $j_{3}$ is different from (\ref{4.33}), but again the difference
vanishes after factor extraction. - $C_{2}(t)$ is, in \cite{vzs},\ the
quantity $G_{11}$\ called the propagator. For the calculations described in
this paper, the formula \cite{vzs}\ (34) is approximated to lowest order,
which in the time domain leads to the equation:%
\begin{equation}
\partial _{t}(C_{2})_{12}(t)=-\omega
_{13}(C_{2})_{32}(t)-\int_{0}^{t}dt^{\prime }(\Gamma _{11})_{13}(t-t^{\prime
})(C_{2})_{32}(t^{\prime })  \label{6.17}
\end{equation}%
\begin{equation}
\omega _{12}=-\langle \dot{h}_{1}h_{2}^{\ast }\rangle  \label{6.18}
\end{equation}

$\Gamma _{11}$ is localized in time, as in (\ref{4.7}). But then $\partial
_{t}C_{2}$\ is discontinuous at $t=0$\ since $\partial
_{t}(C_{2})_{12}(0)=-\omega _{12}$; therefore, the formula resulting from (%
\ref{6.17}) must be written:%
\begin{equation}
\left. 
\begin{array}{c}
\partial _{t}(C_{2})_{12}(t)=-\kappa _{13}(t)(C_{2})_{32}(t) \\ 
\kappa _{13}(t)=\omega _{13}+\theta (t)\gamma _{13} \\ 
\theta (t)=\left\{ 
\begin{array}{c}
1,\;t>0 \\ 
0,\;t\leq 0%
\end{array}%
\right.%
\end{array}%
\right\}  \label{6.20}
\end{equation}

The time dependence of $\kappa _{13}$\ is important when one needs to
calculate the second derivative of $C_{2}$; otherwise, it can be neglected.
We have:%
\begin{equation}
\partial _{t}\kappa _{13}(t)=\delta (t)\gamma _{13}  \label{6.21}
\end{equation}%
\begin{equation}
\partial _{tt}(C_{2})_{12}(t)=-\delta (t)\gamma _{13}(j_{2})_{32}+\kappa
_{13}\kappa _{34}(C_{2})_{42}(t)  \label{6.22}
\end{equation}

In (\ref{6.22}) the denotation $(C_{2})_{12}(0)=\langle h_{1}h_{2}^{\ast
}\rangle =(j_{2})_{12}$ has been used.

\section{Explicit formulation}

\setcounter{equation}{0}

In order to keep the formulation reasonably short, in the past sections of
this paper, a rather formal description has been used employing number
indices; double indizes indicate a component summation and a wave number
integration, together with a factor $(2\pi )^{-3}$. While this is reasonable
for formal calculations, for the rest of this paper we will switch to an
easily readable form where numbers are component indices, and the wave
numbers are explicitly shown. For all correlation functions, we will extract
the factor described in (\ref{4.0.2}), denoting the remaining part with the
same symbol as before. It then becomes apparent that the additional wave
number integrals introduced via the definition of the projection operators (%
\ref{4.1}), (\ref{4.2}), all can be executed. The formula for $\mathbf{D}%
_{3} $, (\ref{5.6}),\ then reads:

\begin{align}
(D_{3})& _{0}(\mathbf{k},t)=  \notag \\
& -\,\frac{1}{2}\rho ^{2}\beta \left( \frac{1}{(2\pi )^{6}}\int d\mathbf{k}%
^{\prime }d\mathbf{k}^{\prime \prime }d\mathbf{k}^{\prime \prime \prime }%
\mathbf{u}_{1}(\mathbf{k}^{\prime },t)\mathbf{u}_{2}(\mathbf{k}^{\prime
\prime },t)\mathbf{u}_{3}(\mathbf{k}^{\prime \prime \prime },t)\,\delta (%
\mathbf{k-\mathbf{k}^{\prime }-\mathbf{k}^{\prime \prime }-k}^{\prime \prime
\prime })\times \right.  \notag \\
& \text{ \ \ \ \ \ \ \ \ \ \ \ \ \ \ \ \ \ \ \ \ \ \ \ \ \ \ \ \ \ \ \ \ \ \
\ \ \ \ \ \ \ \ \ \ \ \ \ \ \ \ \ \ }\times (\Delta _{3})_{0123}(\mathbf{%
\mathbf{k}}^{\prime }\mathbf{,\mathbf{k}^{\prime \prime },k}^{\prime \prime
\prime })  \notag \\
& +\left. \xi ^{-\frac{1}{2}}\frac{1}{(2\pi )^{3}}\int d\mathbf{k}^{\prime }d%
\mathbf{k}^{\prime \prime }u_{1}(\mathbf{\mathbf{k}^{\prime },}t)uq(\mathbf{%
\mathbf{k}^{\prime \prime }}t)\,\delta (\mathbf{k-\mathbf{k}^{\prime }-%
\mathbf{k}^{\prime \prime }})(\Delta _{2})_{01}(\mathbf{\mathbf{k}}^{\prime }%
\mathbf{,\mathbf{k}^{\prime \prime }})\right)  \label{7.1}
\end{align}%
\begin{equation}
(\Delta _{3})_{0123}(\mathbf{\mathbf{k}}^{\prime }\mathbf{,\mathbf{k}%
^{\prime \prime },k}^{\prime \prime \prime })=\int_{0}^{\infty }dt^{\prime
}\Phi 1_{0123}(\mathbf{\mathbf{k}}^{\prime }\mathbf{,\mathbf{k}^{\prime
\prime },k}^{\prime \prime \prime },t^{\prime })+\lim_{t\rightarrow \infty
}\int_{0}^{t}dt^{\prime }\int_{0}^{t-t^{\prime }}dt^{\prime \prime }\Phi
2_{0123}(\mathbf{\mathbf{k}^{\prime },\mathbf{k}^{\prime \prime },k}^{\prime
\prime \prime },t^{\prime },t^{\prime \prime })  \label{7.2}
\end{equation}%
\begin{equation}
(\Delta _{2})_{01}(\mathbf{\mathbf{k}}^{\prime }\mathbf{,\mathbf{k}^{\prime
\prime }})=\int_{0}^{\infty }dt^{\prime }(K_{2})_{01}\left( \mathbf{\mathbf{k%
}^{\prime },\mathbf{k}^{\prime \prime },}t^{\prime }\right)  \label{7.3}
\end{equation}%
\begin{equation}
\xi =\rho \beta \partial _{\rho }P|_{\beta }  \label{7.4}
\end{equation}

The factor $(2\pi )^{3}\delta (\mathbf{k-\mathbf{k}^{\prime }-\mathbf{k}%
^{\prime \prime }-k}^{\prime \prime \prime })$ \ has been extracted from the
quantities \ $\Phi 1$, $\Phi 2$\ in (\ref{a.45});\ also, $(2\pi )^{3}\delta (%
\mathbf{k-\mathbf{k}^{\prime }-\mathbf{k}^{\prime \prime }})$\ has been
extracted from $K_{2}$\ in (\ref{a.46}). \ $uq(\mathbf{\mathbf{k}})$\ is the
Fourier transform of $(\mathbf{u}(x))^{2}$. The resulting explicit formulas
for \ $\Phi 1$, \ $\Phi 2$, $K2$\ are given in the appendix, (\ref{a.47}) to
(\ref{a.49}). - The equation for $C_{2}$\ (\ref{6.20}) now reads:%
\begin{equation}
\partial _{t}(C_{2})_{12}(\mathbf{\mathbf{k},}t)=-\kappa _{13}(\mathbf{%
\mathbf{k},}t)(C_{2})_{32}(\mathbf{\mathbf{k},}t)  \label{7.5}
\end{equation}

We have, after factor extraction, $(j_{2})_{12}=\delta _{12}$ . Therefore,
for the second derivative, one obtains from (\ref{6.22}):

\begin{equation}
\partial _{tt}(C_{2})_{12}(\mathbf{\mathbf{k,}}t)=-\delta (t)\gamma _{12}(%
\mathbf{\mathbf{k}})+\kappa _{13}(\mathbf{\mathbf{k}})\kappa _{34}(\mathbf{%
\mathbf{k}})(C_{2})_{42}(\mathbf{\mathbf{k,}}t)  \label{7.6}
\end{equation}

Instead of (\ref{6.7}), (\ref{6.8}) we have:

\begin{equation}
(C_{3})_{123}(\mathbf{\mathbf{k},\mathbf{k}^{\prime },\mathbf{\mathbf{k}}%
^{\prime \prime },}t)=(C_{2})_{14}(\mathbf{\mathbf{k},}%
t)(j_{3})_{423}+(G_{12})_{123}(\mathbf{\mathbf{k},\mathbf{k}^{\prime },%
\mathbf{\mathbf{k}}^{\prime \prime },}t)  \label{7.7}
\end{equation}%
\begin{equation}
(G_{12})_{123}(\mathbf{\mathbf{k},\mathbf{k}^{\prime },\mathbf{\mathbf{k}}%
^{\prime \prime },}t)=-\int_{0}^{t}dt^{\prime }\,(C_{2})_{14}(\mathbf{%
\mathbf{k,}}t-t^{\prime })(s_{3})_{456}(\mathbf{\mathbf{k}})(C_{2})_{25}(%
\mathbf{\mathbf{k}^{\prime },}t^{\prime })(C_{2})_{36}(\mathbf{\mathbf{k}}%
^{\prime \prime }\mathbf{,}t^{\prime })  \label{7.8}
\end{equation}

We write $C_{3}$\ and $G_{12}$\ as functions of three wave numbers which
correspond to the wave numbers of the three $h$-factors in the definition
formula of $C_{3}$; though, because of $\mathbf{\mathbf{k}}^{\prime \prime }=%
\mathbf{\mathbf{k}}-\mathbf{\mathbf{k}^{\prime }}$,\ they are actually
functions of \ $\mathbf{\mathbf{k}}$, $\mathbf{\mathbf{k}^{\prime }}$\ only.
Corresponding arguments apply for $C_{4}$, $G_{13}$, $\zeta _{3}$, $\zeta
_{4}$. - Finally, (\ref{6.12}) , (\ref{6.13}) turn to:

\begin{align}
(C_{4})_{1234}(\mathbf{\mathbf{k},\mathbf{k}^{\prime },\mathbf{k}^{\prime
\prime },\mathbf{k}}^{\prime \prime \prime }\mathbf{,}t)& =((C_{3})_{156}(%
\mathbf{\mathbf{k},\mathbf{k}^{\prime },\mathbf{k}^{\prime \prime },}%
t)-(j_{3})_{156})\Phi _{23456}+  \notag \\
& +(C_{2})_{17}(\mathbf{\mathbf{k},}t)(j_{4})_{7234}+(G_{13})_{1234}(\mathbf{%
\mathbf{k},\mathbf{k}^{\prime },\mathbf{k}^{\prime \prime },\mathbf{k}}%
^{\prime \prime \prime }\mathbf{,}t)  \label{7.9}
\end{align}

\begin{align}
& (G_{13})_{1234}((\mathbf{\mathbf{k},\mathbf{k}^{\prime },\mathbf{k}%
^{\prime \prime },\mathbf{k}}^{\prime \prime \prime }\mathbf{,}t)=  \notag \\
& =-\int_{0}^{t}dt^{\prime }\,(C_{2})_{15}(\mathbf{\mathbf{k},}t-t^{\prime
})(\sigma _{4})_{5678}(\mathbf{\mathbf{k}})(C_{2})_{26}(\mathbf{\mathbf{k}%
^{\prime },}t^{\prime })(C_{2})_{37}(\mathbf{\mathbf{k}^{\prime \prime },}%
t^{\prime })(C_{2})_{48}((\mathbf{\mathbf{k}}^{\prime \prime \prime }\mathbf{%
,}t^{\prime })  \label{7.10}
\end{align}

with $\mathbf{\mathbf{k}}=\mathbf{\mathbf{k}^{\prime }+\mathbf{k}^{\prime
\prime }+\mathbf{k}}^{\prime \prime \prime }$.

\section{Substitution of time derivatives; time integration}

\setcounter{equation}{0}

The results of sec. 6, in the form given in sec. 7, have to be introduced
into the formulas for $\Phi 1$, $\Phi 2$, $K_{2}$. Here, time derivatives of 
$C_{3}$, $C_{4}$ up to the second order appear. From (\ref{7.7}), we have,
with (\ref{7.5}), (\ref{7.6}):

\begin{equation}
\partial _{t}(C_{3})_{123}(\mathbf{\mathbf{k},\mathbf{k}^{\prime },\mathbf{k}%
}^{\prime \prime }\mathbf{,}t)=-\kappa _{15}(\mathbf{\mathbf{k},}%
t)(C_{2})_{54}(\mathbf{\mathbf{k},}t)(j_{3})_{423}+\partial
_{t}(G_{12})_{123}(\mathbf{\mathbf{k},\mathbf{k}^{\prime },\mathbf{k}}%
^{\prime \prime }\mathbf{,}t)  \label{8.1}
\end{equation}

\begin{align}
\partial _{tt}& (C_{3})_{123}(\mathbf{\mathbf{k},\mathbf{k}^{\prime },%
\mathbf{k}}^{\prime \prime }\mathbf{,}t)=  \notag \\
& =(-\delta (t)\gamma _{15}(\mathbf{\mathbf{k}})+\kappa _{16}(\mathbf{%
\mathbf{k}})\kappa _{65}(\mathbf{\mathbf{k}})(C_{2})_{54}(\mathbf{\mathbf{k,}%
}t))(j_{3})_{423}+\partial _{tt}(G_{12})_{123}(\mathbf{\mathbf{k},\mathbf{k}%
^{\prime },\mathbf{k}}^{\prime \prime }\mathbf{,}t)  \label{8.2}
\end{align}

From (\ref{7.8}), we obtain:%
\begin{align}
\partial _{t}& (G_{12})_{123}(\mathbf{\mathbf{k},\mathbf{k}^{\prime },%
\mathbf{k}}^{\prime \prime }\mathbf{,}t)=-(s_{3})_{156}(\mathbf{\mathbf{k}}%
)(C_{2})_{25}(\mathbf{\mathbf{k}^{\prime },}t)(C_{2})_{36}(\mathbf{\mathbf{k}%
}^{\prime \prime }\mathbf{,}t)+  \notag \\
& +\int_{0}^{t}dt^{\prime }\,\kappa _{17}(\mathbf{\mathbf{k},}t-t^{\prime
})(C_{2})_{74}(\mathbf{\mathbf{k,}}t-t^{\prime })(s_{3})_{456}(\mathbf{%
\mathbf{k}})(C_{2})_{25}(\mathbf{\mathbf{k}^{\prime },}t^{\prime
})(C_{2})_{36}(\mathbf{\mathbf{k}}^{\prime \prime }\mathbf{,}t^{\prime }) 
\notag \\
& =-(s_{3})_{156}(\mathbf{\mathbf{k}})(C_{2})_{25}(\mathbf{\mathbf{k}%
^{\prime },}t)(C_{2})_{36}(\mathbf{\mathbf{k}}^{\prime \prime }\mathbf{,}%
t)-\kappa _{14}(\mathbf{\mathbf{k}})(G_{12})_{423}(\mathbf{\mathbf{k},%
\mathbf{k}^{\prime },\mathbf{k}}^{\prime \prime }\mathbf{,}t)  \label{8.3}
\end{align}

We remember that $\mathbf{\mathbf{k}}^{\prime \prime }=\mathbf{\mathbf{k}}-%
\mathbf{\mathbf{k}^{\prime }}$. The first step is for calculation of the
second-order derivative, where it is essential that $\kappa $\ depends on
time; the second step is for insertion into the main formula. For $\partial
_{tt}G_{12}$ we obtain, after some manipulations:%
\begin{eqnarray}
\partial _{tt}(G_{12})_{123}(\mathbf{\mathbf{k},\mathbf{k}^{\prime },\mathbf{%
k}}^{\prime \prime }\mathbf{,}t) &=&(\zeta _{3})_{145678}(\mathbf{\mathbf{k},%
\mathbf{k}^{\prime },\mathbf{k}}^{\prime \prime })(s_{3})_{645}(\mathbf{%
\mathbf{k}})(C_{2})_{27}(\mathbf{\mathbf{k}^{\prime },}t)(C_{2})_{38}(%
\mathbf{\mathbf{k}}^{\prime \prime }\mathbf{,}t)+  \notag \\
&&\text{ \ \ \ \ \ \ \ \ \ \ \ \ \ \ \ \ \ \ \ \ \ \ \ }+\kappa _{14}(%
\mathbf{\mathbf{k}})\kappa _{45}(\mathbf{\mathbf{k}})(G_{12})_{523}(\mathbf{%
\mathbf{k},\mathbf{k}^{\prime },\mathbf{k}}^{\prime \prime }\mathbf{,}t)
\label{8.4}
\end{eqnarray}%
\begin{equation}
(\zeta _{3})_{123456}(\mathbf{\mathbf{k},\mathbf{k}^{\prime },\mathbf{k}}%
^{\prime \prime })=\kappa _{14}(\mathbf{\mathbf{k}})\delta _{25}\delta
_{36}+\delta _{14}\kappa _{25}(\mathbf{\mathbf{k}}^{\prime })\delta
_{36}+\delta _{14}\delta _{25}\kappa _{36}(\mathbf{\mathbf{k}}^{\prime
\prime })  \label{8.5}
\end{equation}

By a similar calculation, we obtain for $C_{4}$, from (\ref{7.9}):

\begin{align}
\partial _{t}(C_{4})_{1234}(\mathbf{\mathbf{k},\mathbf{k}^{\prime },\mathbf{k%
}^{\prime \prime },}t)& =\partial _{t}(C_{3})_{156}(\mathbf{\mathbf{k},%
\mathbf{k}^{\prime },}t)\Phi _{23456}+  \notag \\
& +\partial _{t}(C_{2})_{17}(\mathbf{\mathbf{k},}t)(j_{4})_{7234}+\partial
_{t}(G_{13})_{1234}(\mathbf{\mathbf{k},\mathbf{k}^{\prime },\mathbf{k}%
^{\prime \prime },}t)  \label{8.6}
\end{align}

\begin{align}
\partial _{tt}(C_{4})_{1234}(\mathbf{\mathbf{k},\mathbf{k}^{\prime },\mathbf{%
k}^{\prime \prime },}t)& =\partial _{tt}(C_{3})_{156}(\mathbf{\mathbf{k},%
\mathbf{k}^{\prime },}t)\Phi _{23456}+  \notag \\
& +\partial _{tt}(C_{2})_{17}(\mathbf{\mathbf{k},}t)(j_{4})_{7234}+\partial
_{tt}(G_{13})_{1234}(\mathbf{\mathbf{k},\mathbf{k}^{\prime },\mathbf{k}%
^{\prime \prime },}t)  \label{8.7}
\end{align}

For the time derivatives of $C_{2}$\ and $C_{3}$, (\ref{7.5}), (\ref{7.6}), (%
\ref{8.1}), (\ref{8.2}) can be used for substitution. From (\ref{7.10}), we
find:%
\begin{align}
& \partial _{t}(G_{13})_{1234}((\mathbf{\mathbf{k},\mathbf{k}^{\prime },%
\mathbf{k}^{\prime \prime },\mathbf{k}}^{\prime \prime \prime },t)=  \notag
\\
& =-(\sigma _{4})_{1567}(\mathbf{\mathbf{k}})(C_{2})_{25}(\mathbf{\mathbf{k}%
^{\prime },}t)(C_{2})_{36}(\mathbf{\mathbf{k}^{\prime \prime },}%
t)(C_{2})_{47}((\mathbf{\mathbf{k}}^{\prime \prime \prime }\mathbf{,}%
t)-\kappa _{15}(\mathbf{\mathbf{k}})(G_{13})_{5234}((\mathbf{\mathbf{k},%
\mathbf{k}^{\prime },\mathbf{k}^{\prime \prime },\mathbf{k}}^{\prime \prime
\prime },t)  \label{8.8}
\end{align}%
\begin{align}
& \partial _{tt}(G_{13})_{1234}((\mathbf{\mathbf{k},\mathbf{k}^{\prime },%
\mathbf{k}^{\prime \prime },\mathbf{k}}^{\prime \prime \prime },t)=  \notag
\\
& =(\zeta _{4})_{15678\bar{2}\bar{3}\bar{4}}(\mathbf{\mathbf{k},\mathbf{k}%
^{\prime },\mathbf{k}}^{\prime \prime }\mathbf{,\mathbf{k}}^{\prime \prime
\prime })(\sigma _{4})_{8567}(\mathbf{\mathbf{k}})(C_{2})_{2\bar{2}}(\mathbf{%
\mathbf{k}^{\prime },}t)(C_{2})_{3\bar{3}}(\mathbf{\mathbf{k}^{\prime \prime
},}t)(C_{2})_{4\bar{4}}((\mathbf{\mathbf{k}}^{\prime \prime \prime }\mathbf{,%
}t)+  \notag \\
& \text{ \ \ \ \ \ \ \ \ \ \ \ \ \ \ \ \ \ \ \ \ \ \ \ \ \ \ \ \ \ \ \ \ }%
+\kappa _{15}(\mathbf{\mathbf{k}})\kappa _{56}(\mathbf{\mathbf{k}}%
)(G_{13})_{6234}((\mathbf{\mathbf{k},\mathbf{k}^{\prime },\mathbf{k}^{\prime
\prime },\mathbf{k}}^{\prime \prime \prime },t)  \label{8.9}
\end{align}%
\begin{align}
(\zeta _{4})& _{12345678}(\mathbf{\mathbf{k},\mathbf{k}^{\prime },\mathbf{k}}%
^{\prime \prime }\mathbf{,\mathbf{k}}^{\prime \prime \prime })=  \notag \\
& =\kappa _{15}(\mathbf{\mathbf{k}})\delta _{26}\delta _{37}\delta
_{48}+\delta _{15}\kappa _{26}(\mathbf{\mathbf{k}}^{\prime })\delta
_{37}\delta _{48}+\delta _{15}\delta _{26}\kappa _{37}(\mathbf{\mathbf{k}}%
^{\prime \prime })\delta _{48}+\delta _{15}\delta _{26}\delta _{37}\kappa
_{48}(\mathbf{\mathbf{k}}^{\prime \prime \prime })  \label{8.10}
\end{align}

The formulas derived in this section are inserted into $\Phi 1$, $\Phi 2$.
It is seen that terms still containing $G_{12}$, $G_{13}$ cancel each other.
The remaining terms are products of $C_{2}$\ of different order (in $\Phi 2$%
\ with different time variables) multiplied by certain coefficients. Now we
are able to perform the time integrations in (\ref{7.2}). The solution of (%
\ref{7.5}) can be expressed by the matrix exponential function. The time
dependence of $\kappa $\ can now be neglected:%
\begin{equation}
(C_{2})_{12}(\mathbf{\mathbf{k},}t)=\func{e}^{-\kappa _{12}(\mathbf{\mathbf{%
k)}}t}  \label{8.11}
\end{equation}

The integrations in (\ref{7.2}) result in inverses of $\zeta _{3}$, $\zeta
_{4}$\ and of:%
\begin{equation}
(\zeta _{2})_{1234}(\mathbf{\mathbf{k},\mathbf{k}^{\prime }})=\kappa _{13}(%
\mathbf{\mathbf{k}})\delta _{24}+\delta _{13}\kappa _{24}(\mathbf{\mathbf{k}}%
^{\prime })  \label{8.12}
\end{equation}%
\begin{equation}
(\check{\zeta}_{3})_{123456}(\mathbf{\mathbf{k},\mathbf{k}^{\prime },\mathbf{%
k}^{\prime \prime }})=-\kappa _{14}(\mathbf{\mathbf{k}})\delta _{25}\delta
_{36}+\delta _{14}\kappa _{25}(\mathbf{\mathbf{k}}^{\prime })\delta
_{36}+\delta _{14}\delta _{25}\kappa _{36}(\mathbf{\mathbf{k}}^{\prime
\prime })  \label{8.13}
\end{equation}

For instance, we have:%
\begin{equation}
\int_{0}^{\infty }dt(C_{2})_{12}(\mathbf{\mathbf{k},}t)(C_{2})_{34}(\mathbf{%
\mathbf{k}}^{\prime }\mathbf{,}t)=(\zeta _{2})_{1324}^{-1}(\mathbf{\mathbf{k}%
,\mathbf{k}^{\prime }})  \label{8.14}
\end{equation}

After integration, the parts from $\Phi 1$\ and $\Phi 2$\ can be united.
Some terms cancel, others can be simplified. For instance, the relation
applies:%
\begin{equation}
(\zeta _{3})_{123478}(\mathbf{\mathbf{k},\mathbf{k}^{\prime },\mathbf{k}%
^{\prime \prime }})(\zeta _{2})_{7856}^{-1}(\mathbf{\mathbf{k}}^{\prime }%
\mathbf{,\mathbf{k}^{\prime \prime }})=\kappa _{14}(\mathbf{\mathbf{k}}%
)(\zeta _{2})_{2356}^{-1}(\mathbf{\mathbf{k}}^{\prime }\mathbf{,\mathbf{k}%
^{\prime \prime }})+\delta _{14}\delta _{25}\delta _{36}  \label{8.15}
\end{equation}

From this calculation, we obtain an expression for the kernel function $%
\Delta _{3}$\ (\ref{7.2}) which is presented in the appendix, (\ref{a.50}).
After combining some similar quantities, it consists of 10 terms. A problem
arises with terms no. 7 and 9 where wave number integrals appear which
contain static correlations. Within the integral, all external wave numbers
have been set equal to zero; the appearing wave number arguments pertain to
the indices 5 and 6. Here it is not possible to localize the integrand
factors totally since then the integral will diverge. We will discuss this
in the next section.

\section{Modes; static correlations; 2$^{nd}$ order term}

\setcounter{equation}{0}

During the investigation, I usually denoted a component index which runs
from 1 ot 5 with a greek symbol; and one running over 3 values with a latin.
In order to have a short notation, I sometimes speak of greek and latin
indices; and, describing an indexed quantity, I call it greek or latin, or I
speak of its greec or latin values. - Component indices generally are greek;
but indices 0 to 3 in (\ref{7.1}), are latin. - To proceed further, we make
use of the mode analysis of hydrodynamics. The (greek) matrix $\kappa $\ is
identical to the matrix of linearized hydrodynamics; see, e. g., \cite{ehl}.
It is symmetric in its indices and therefore can be decomposed into
hydrodynamic modes $\hat{\kappa}$:%
\begin{equation}
\kappa _{12}=\chi _{13}\chi _{23}\hat{\kappa}_{3}  \label{9.1}
\end{equation}

$\chi $\ is the modal matrix which facilitates the decomposition. The modes
are: two sound modes, a heat mode and two identical shear modes. Since in
this investigation we are not interested in sound and heat effects, we
neglect all parts of $\kappa $\ connected to the first three modes. Then $%
\kappa $\ simplifies considerably:\ It now is latin, and it reads:%
\begin{equation}
\begin{array}{c}
\kappa _{12}(\mathbf{\mathbf{k}})=-\varepsilon _{12}(\mathbf{\mathbf{k}}%
)\,\nu \\ 
\varepsilon _{12}(\mathbf{\mathbf{k}})=k^{2}\delta _{12}-k_{1}k_{2}%
\end{array}
\label{9.2}
\end{equation}

$\nu $ is the kinematic shear viscosity. - Now that $\kappa $ is latin, this
also applies for matrices $\zeta _{2}$, $\zeta _{3}$, $\check{\zeta}_{3}$, $%
\zeta _{4}$\ and their inverses. One then finds that all component indices
of (\ref{a.50}) are latin, with the exception of indices 5, 6 in the wave
number integrals in terms 7 and 9.

Static correlations can be calculated with a method described in \cite{mu}.
A modified equilibrium probability density is used which describes a fluid
which moves with a uniform velocity $\mathbf{u}$. Expectations of the
conserved quantity densities and their fluxes are formulated;\ derivatives
of several order with respect to the the conjugate parameters are taken; it
is shown that the resulting formulas for $\mathbf{u=0}$\ are the static
correlations.

Many elements of the static correlation matrices are zero. As a first step
of application of the described technique, one needs a means for finding
these zero elements. The rule is this: Let $n$ be the order of the static
correlation. Let $\eta $\ be the number of (generally greek) indices whose
values are not latin. Build $n+\eta .$Write $e$ for the case that this
number is even and $o$ \ that it is odd. Write, for short, $j_{n}$\ and $%
s_{n}$\ for the two types of static correlations which appear in the
formulas. Then:%
\begin{equation}
\begin{tabular}{c|cc}
$n+\eta $ & $e$ & $o$ \\ \hline
$j_{n}$ & $\neq 0$ & $0$ \\ 
$s_{n}$ & $0$ & $\neq 0$%
\end{tabular}
\label{9.3}
\end{equation}

$\neq 0$ means 'generally not zero'. For the majority of coefficients in (%
\ref{a.50}) all indices are latin, so $\eta =0$. We find that the latin
elements of $s_{4}$\ and $j_{3}$\ are zero; so, in (\ref{a.50}), the terms 1
an 4 as well as the second parts of terms 7 and 9 are zero. The wave number
integrals in terms 7 and 9 remain being a problem. We will assume here that
they are somehow related to the localized value of the integrand (i. e. the
integrand with all wave numbers equal to zero). In term 9, all indices of $%
\Phi $\ are latin. With $j_{5}$\ and $j_{3}$\ equal to zero, $\Phi $\ is
zero too. In term 7, indices 4, 7, 8 are latin while 5, 6 are either both
latin or both non-latin (1 oder 2), since otherwise the second factor of the
integrand will be zero. In the first case, $\Phi =0$. In the second case, in
(\ref{6.14}), the second term is zero; in the first term, for $j_{5}$, $\eta
=2$\ , and for the third term, for $j_{3},\eta =2$\ also;\ then, from (\ref%
{9.3}) both coefficients are zero. Therefore, in the wave number integrals
in terms 7 and 9, the localized value of the integrand is zero. This does
not necessarily include that the integrals vanish. But, with this argument,
we will neglect them. - The remaining formula for $\Delta _{3}$\ is given in
the appendix (\ref{a.51}).

The non-zero static correlation components are calculated with the technique
outlined above. One obtains (all indices latin):%
\begin{equation}
(j_{4})_{1234}=\xi ^{-1}(\delta _{12}\delta _{34}+\delta _{13}\delta
_{24}+\delta _{14}\delta _{23})  \label{9.4}
\end{equation}%
\begin{equation}
(s_{3})_{123}(\mathbf{\mathbf{k}})=ik_{4}\frac{1}{\sqrt{\rho \beta }}(\delta
_{12}\delta _{34}+\delta _{13}\delta _{24}-\lambda \delta _{14}\delta _{23})
\label{9.5}
\end{equation}%
\begin{equation}
\lambda =\frac{\partial _{\beta }p|_{\rho }}{\partial _{\beta }\varepsilon
|_{\rho }}  \label{9.6}
\end{equation}

$p$ is the static pressure and $\varepsilon $\ the energy density, both as
functions of mass density $\rho $\ and inverse kinetic temperature $\beta $.
Via $\lambda $, the formulas are dependent on the physical properties of the
fluid.

We will show that the second-order term (the last row of (\ref{7.1})) just
cancels the first term of \ (\ref{a.51}). The part of $D_{3}$ which contains
this term simplifies somewhat when (\ref{9.4})\ is introduced:

\begin{align}
& -\,\frac{1}{2}\rho ^{2}\beta \frac{1}{(2\pi )^{6}}\int d\mathbf{k}^{\prime
}d\mathbf{k}^{\prime \prime }d\mathbf{k}^{\prime \prime \prime }\mathbf{u}%
_{1}(\mathbf{k}^{\prime },t)\mathbf{u}_{4}(\mathbf{k}^{\prime \prime },t)%
\mathbf{u}_{4}(\mathbf{k}^{\prime \prime \prime },t)\,\delta (\mathbf{k-%
\mathbf{k}^{\prime }-\mathbf{k}^{\prime \prime }-k}^{\prime \prime \prime
})\times  \notag \\
& \text{ \ \ \ \ \ \ \ \ \ \ \ \ \ \ \ \ \ \ \ \ \ \ \ \ \ \ \ \ \ \ \ \ \ \
\ \ \ \ \ \ \ \ \ \ \ \ \ \ \ \ \ \ }\times \gamma _{01}(\mathbf{k})(\xi
^{-1}-(2\pi )^{3}\delta (\mathbf{k}-\mathbf{k}^{\prime }))  \label{9.7}
\end{align}

The qunatity $K_{2}$\ (\ref{a.49}) has been processed by introducing the
formulas for $C_{3}$\ (\ref{6.7}), (\ref{6.8}) and its derivatives (\ref{8.1}%
) to (\ref{8.4}). Then, $G_{12}$\ does not explicitly show up any more, and $%
K_{2}$\ now reads:

\begin{align}
(& K_{2})_{01}(\mathbf{\mathbf{k}^{\prime },\mathbf{k}}^{\prime \prime
},t)=(\gamma _{01}(\mathbf{k})(2\pi )^{3}\delta (\mathbf{k}^{\prime \prime
})\langle h_{2}\rangle -\gamma _{04}(\mathbf{k})\,j_{412})\delta (t)  \notag
\\
& -i\,k_{6}^{\prime \prime }{}(C_{2})_{56}(\mathbf{k}^{\prime \prime
},t)(C_{2})_{41}(\mathbf{k}^{\prime },t)s_{045}(\mathbf{k})  \notag \\
& +(C_{2})_{52}(\mathbf{k}^{\prime \prime },t)\left( -(C_{2})_{43}(\mathbf{k}%
^{\prime },t)s_{045}(\mathbf{k})\gamma _{31}(\mathbf{k}^{\prime
})+(C_{2})_{41}(\mathbf{k}^{\prime },t)s_{067}(\mathbf{k})(\zeta
_{2})_{6745}((\mathbf{\mathbf{k}^{\prime },\mathbf{k}}^{\prime \prime
})\right)  \label{9.8}
\end{align}

Remember that the component index 2 has the fixed value 1. The formula now
consists of 3 terms. The second term, after time integration, contains the
factor $k_{6}^{\prime \prime }{}(\zeta _{2})_{4156}^{-1}(\mathbf{\mathbf{k}%
^{\prime },\mathbf{k}}^{\prime \prime })$ which is zero. The third term
shows a $C_{2}$\ factor one index of which has the fixed value 1. When the
analysis is restricted to shear modes, as is introduced in the previous
section, this term does not contribute. Thus, in (\ref{9.8}), the first term
remains. - We can write $h_{2}=h_{1}$\ , where on the left the index
variable 2 appears and on the right its fixed value 1;\ we have $\langle
h_{1}\rangle =\xi ^{\frac{1}{2}}$. For the coefficient of the second partial
term we find $\,j_{412}=\xi ^{-\frac{1}{2}}\delta _{41}$. The last row of (%
\ref{7.1}) now reads:\ 

\begin{equation}
\,\frac{1}{2}\rho ^{2}\beta \frac{1}{(2\pi )^{3}}\int d\mathbf{k}^{\prime }d%
\mathbf{k}^{\prime \prime }u_{1}(\mathbf{\mathbf{k}^{\prime },}t)uq(\mathbf{%
\mathbf{k}^{\prime \prime }}t)\,\delta (\mathbf{k-\mathbf{k}^{\prime }-%
\mathbf{k}^{\prime \prime }})\gamma _{01}(\mathbf{k})(\xi ^{-1}-(2\pi
)^{3}\delta (\mathbf{k}^{\prime \prime })\,)  \label{9.9}
\end{equation}

The main difference to (\ref{9.7}), apart from the order of integration, is
the appearance of $uq$. The relation to the $\mathbf{u}$'s is:

\begin{eqnarray}
uq(\mathbf{k},t) &=&\int d\mathbf{x}\func{e}^{-i\mathbf{k\cdot x}}u_{4}(%
\mathbf{x},t)u_{4}(\mathbf{x},t)  \notag \\
&=&\frac{1}{(2\pi )^{3}}\int d\mathbf{k}^{\prime }\int d\mathbf{k}^{\prime
\prime }\delta (\mathbf{k-\mathbf{k}}^{\prime }\mathbf{-k}^{\prime \prime
})u_{4}(\mathbf{k}^{\prime },t)u_{4}(\mathbf{k}^{\prime \prime },t)
\label{9.10}
\end{eqnarray}

When this is introduced into (\ref{9.9}) (and some wave numbers are
renamed), it is seen that (\ref{9.9}) actually cancels (\ref{9.7}), as is
stated in the beginning of the section.

\bigskip

\section{Final form of the 3$^{rd}$ order term}

\setcounter{equation}{0}

In (\ref{a.51}), in addition to eliminating the first term, the reduction to
shear modes has to be introduced. For $(\zeta _{2})^{-1}$, $(\check{\zeta}%
_{3})^{-1}$\ we obtain:

\begin{equation}
(\zeta _{2})_{1234}^{-1}(\mathbf{k}^{\prime },\mathbf{k}^{\prime \prime })=%
\frac{\varepsilon _{13}(\mathbf{k}^{\prime })\varepsilon _{24}(\mathbf{k}%
^{\prime \prime })}{k^{\prime 2}k^{\prime \prime 2}\nu (k^{\prime
2}+k^{\prime \prime 2})}  \label{10.1}
\end{equation}

\begin{equation}
(\tilde{\zeta}_{3})_{123456}^{-1}(\mathbf{k}^{\prime },\mathbf{k}^{\prime
\prime },\mathbf{k}^{\prime \prime \prime })=\frac{\varepsilon _{14}(\mathbf{%
k}^{\prime })\varepsilon _{25}(\mathbf{k}^{\prime \prime })\varepsilon _{36}(%
\mathbf{k}^{\prime \prime \prime })}{k^{\prime 2}k^{\prime \prime
2}k^{\prime \prime \prime 2}\nu (-k^{\prime 2}+k^{\prime \prime 2}+k^{\prime
\prime \prime 2})}  \label{10.2}
\end{equation}

When this is inserted into (\ref{a.51}) and then in (\ref{7.1}), after
expansion some terms which stem from the second term of (\ref{9.2}) vanish
because of the continuity condition (\ref{2.14}). Then, it becomes apparent
that the last three terms of (\ref{a.51}) cancel each other, while the two
remaining terms can be somewhat simplified. The final form in wave number
space is:

\begin{align}
(D_{3})& _{0}(\mathbf{k},t)=  \notag \\
\frac{\rho }{\nu }& \frac{1}{(2\pi )^{6}}\int d\mathbf{k}^{\prime }d\mathbf{k%
}^{\prime \prime }d\mathbf{k}^{\prime \prime \prime }\mathbf{u}_{1}(\mathbf{k%
}^{\prime },t)\mathbf{u}_{2}(\mathbf{k}^{\prime \prime },t)\mathbf{u}_{3}(%
\mathbf{k}^{\prime \prime \prime },t)\,\delta (\mathbf{k-\mathbf{k}^{\prime
}-\mathbf{k}^{\prime \prime }-k}^{\prime \prime \prime })\times  \notag \\
& \text{ \ \ \ \ \ \ \ \ \ \ \ \ \ \ \ \ \ \ \ \ \ \ \ \ \ \ \ \ \ \ \ \ \ \
\ \ \ \ \ \ \ \ \ \ \ \ \ \ \ \ \ \ }\times \Delta _{0123}(\mathbf{\mathbf{k}%
}^{\prime }\mathbf{,\mathbf{k}^{\prime \prime },k}^{\prime \prime \prime })
\label{10.3}
\end{align}%
\begin{align}
\Delta & _{0123}(\mathbf{\mathbf{k}}^{\prime }\mathbf{,\mathbf{k}^{\prime
\prime },k}^{\prime \prime \prime })=(s_{3})_{025}(\mathbf{\mathbf{k}}%
^{\prime }+\mathbf{\mathbf{k}^{\prime \prime }+k}^{\prime \prime \prime
})\varepsilon _{56}(\mathbf{k}^{\prime }+\mathbf{k}^{\prime \prime \prime
})\times  \notag \\
& \times \left( -(s_{3})_{136}(\mathbf{k}^{\prime })\frac{1}{(\mathbf{k}%
^{\prime }+\mathbf{k}^{\prime \prime \prime })^{2}(k^{\prime \prime 2}+(%
\mathbf{k}^{\prime }+\mathbf{k}^{\prime \prime \prime })^{2})}+\right. 
\notag \\
& \text{ \ \ \ \ \ \ \ \ \ \ \ \ \ \ \ \ }\left. +(s_{3})_{613}(\mathbf{k}%
^{\prime }+\mathbf{k}^{\prime \prime \prime })\frac{k^{\prime \prime \prime
2}}{(\mathbf{k}^{\prime }+\mathbf{k}^{\prime \prime \prime })^{2}(k^{\prime
\prime 2}+(\mathbf{k}^{\prime }+\mathbf{k}^{\prime \prime \prime
})^{2})(k^{\prime 2}+k^{\prime \prime \prime 2})}\right)  \label{10.4}
\end{align}

We extracted the thermodynamic factors from the $s_{3}$\ coefficients in
order to collect them in front of (\ref{10.3}). We now have:%
\begin{eqnarray}
(s_{3})_{123}(\mathbf{\mathbf{k}}) &=&ik_{4}(s_{3})_{1234}  \notag \\
(s_{3})_{1234} &=&\delta _{12}\delta _{34}+\delta _{13}\delta _{24}-\lambda
\delta _{14}\delta _{23}  \label{10.5}
\end{eqnarray}

The quantites on the left are distinguished by the number of indices. - In
order to transform (\ref{10.3}), (\ref{10.4}) back to geometric space, is
seems reasonable to switch to a different notation:

\begin{align}
(D_{3})& _{0}(\mathbf{k},t)=  \notag \\
=\frac{\rho }{\nu }& \int d\mathbf{k}^{\prime }d\mathbf{k}^{\prime \prime }d%
\mathbf{k}^{\prime \prime \prime }\,\delta (\mathbf{k-\mathbf{k}^{\prime }-%
\mathbf{k}^{\prime \prime }-k}^{\prime \prime \prime })\sum_{\mu
=1}^{2}M_{\mu }(\mathbf{k}^{\prime }\mathbf{,k^{\prime \prime },k}^{\prime
\prime \prime })(Q_{\mu })_{0}(\mathbf{k}^{\prime }\mathbf{,k^{\prime \prime
},k}^{\prime \prime \prime })  \label{10.6}
\end{align}

\begin{eqnarray}
M_{1} &=&\frac{1}{(\mathbf{k}^{\prime }+\mathbf{k}^{\prime \prime \prime
})^{2}(k^{\prime \prime 2}+(\mathbf{k}^{\prime }+\mathbf{k}^{\prime \prime
\prime })^{2})}  \notag \\
M_{2} &=&\frac{1}{(\mathbf{k}^{\prime }+\mathbf{k}^{\prime \prime \prime
})^{2}(k^{\prime \prime 2}+(\mathbf{k}^{\prime }+\mathbf{k}^{\prime \prime
\prime })^{2})(k^{\prime 2}+k^{\prime \prime \prime 2})}  \label{10.7}
\end{eqnarray}%
\begin{equation}
(Q_{1})_{0}=-i(k_{7}^{\prime }+k_{7}^{\prime \prime }+k_{7}^{\prime \prime
\prime })(s_{3})_{0257}(ik_{8}^{\prime })(s_{3})_{1368}\varepsilon _{56}(%
\mathbf{k}^{\prime }+\mathbf{k}^{\prime \prime \prime })\mathbf{u}_{1}(%
\mathbf{k}^{\prime },t)\mathbf{u}_{2}(\mathbf{k}^{\prime \prime },t)\mathbf{u%
}_{3}(\mathbf{k}^{\prime \prime \prime },t)  \label{10.7a}
\end{equation}%
\begin{equation}
(Q_{2})_{0}=i(k_{7}^{\prime }+k_{7}^{\prime \prime }+k_{7}^{\prime \prime
\prime })(s_{3})_{0257}i(k_{8}^{\prime }+k_{8}^{\prime \prime \prime
})(s_{3})_{6138}k^{\prime \prime \prime 2}\varepsilon _{56}(\mathbf{k}%
^{\prime }+\mathbf{k}^{\prime \prime \prime })\mathbf{u}_{1}(\mathbf{k}%
^{\prime },t)\mathbf{u}_{2}(\mathbf{k}^{\prime \prime },t)\mathbf{u}_{3}(%
\mathbf{k}^{\prime \prime \prime },t)  \label{10.8}
\end{equation}

Then, the formal expression for the transformed quantity is:

\begin{align}
(D_{3})& _{0}(\mathbf{x},t)=  \notag \\
=\frac{\rho }{\nu }& \int d\mathbf{x}^{\prime }d\mathbf{x}^{\prime \prime }d%
\mathbf{x}^{\prime \prime \prime }\,\sum_{\mu =1}^{2}M_{\mu }(\mathbf{x-x}%
^{\prime }\mathbf{,x-x^{\prime \prime },x-x}^{\prime \prime \prime })(Q_{\mu
})_{0}(\mathbf{x}^{\prime }\mathbf{,x^{\prime \prime },x}^{\prime \prime
\prime })  \label{10.9}
\end{align}

The functions in this formula are the (backwards) Fourier Transforms of the
equally denoted functions in (\ref{10.6}). The transform of the kernel
functions (\ref{10.7}) can be found in closed form:%
\begin{eqnarray}
M_{1}(\mathbf{x}^{\prime }\mathbf{,x^{\prime \prime },x}^{\prime \prime
\prime }) &=&\frac{1}{2^{4}\pi ^{3}x^{\prime }x^{\prime \prime }}\arctan (%
\frac{x^{\prime }}{x^{\prime \prime }})\delta (\mathbf{x}^{\prime }-\mathbf{x%
}^{\prime \prime \prime })  \notag \\
M_{2}(\mathbf{x}^{\prime }\mathbf{,x^{\prime \prime },x}^{\prime \prime
\prime }) &=&\frac{1}{2^{8}\pi ^{4}x^{\prime \prime }\tilde{x}x^{\ast }}%
\arctan (\frac{\tilde{x}}{\hat{x}})  \label{10.10}
\end{eqnarray}%
\begin{eqnarray}
\mathbf{\tilde{x}} &\mathbf{=}&\frac{1}{2}(\mathbf{x}^{\prime }\mathbf{-x}%
^{\prime \prime \prime })  \notag \\
\mathbf{x}^{\ast } &\mathbf{=}&\frac{1}{2}(\mathbf{x}^{\prime }+\mathbf{x}%
^{\prime \prime \prime })  \notag \\
\hat{x} &=&x^{\prime \prime }+x^{\ast }  \label{10.11}
\end{eqnarray}

As usual, vectors are in bold letters, while their magnitude is in normal
letters. The velocity functions (\ref{10.8}) are transformed
straightforwardly:%
\begin{equation}
(Q_{1})_{0}=(s_{3})_{0257}(s_{3})_{1368}(\nabla _{7}^{\prime }+\nabla
_{7}^{\prime \prime }+\nabla _{7}^{\prime \prime \prime })\nabla
_{8}^{\prime }\varepsilon _{56}(\mathbf{\nabla }^{\prime }+\mathbf{\nabla }%
^{\prime \prime \prime })\mathbf{u}_{1}(\mathbf{x}^{\prime },t)\mathbf{u}%
_{2}(\mathbf{x}^{\prime \prime },t)\mathbf{u}_{3}(\mathbf{x}^{\prime \prime
\prime },t)  \label{10.12}
\end{equation}%
\begin{equation}
(Q_{1})_{0}=(s_{3})_{0257}(s_{3})_{6138}(\nabla _{7}^{\prime }+\nabla
_{7}^{\prime \prime }+\nabla _{7}^{\prime \prime \prime })(\nabla
_{8}^{\prime }+\nabla _{8}^{\prime \prime \prime })\mathbf{\nabla }^{\prime
\prime \prime 2}\varepsilon _{56}(\mathbf{\nabla }^{\prime }+\mathbf{\nabla }%
^{\prime \prime \prime })\mathbf{u}_{1}(\mathbf{x}^{\prime },t)\mathbf{u}%
_{2}(\mathbf{x}^{\prime \prime },t)\mathbf{u}_{3}(\mathbf{x}^{\prime \prime
\prime },t)  \label{10.13}
\end{equation}%
\begin{equation}
\varepsilon _{12}(\mathbf{\nabla })=\mathbf{\nabla }^{2}\delta _{12}-\nabla
_{1}\nabla _{2}  \label{10.14}
\end{equation}

Expansion of (\ref{10.12}), (\ref{10.13}) leads to lenghty expressions which
will not be given here.

\section{Summary and outlook}

\setcounter{equation}{0}

The equation for the hydrodynamic velocity derived by N-particle statistical
mechanics has been investigated. It is argued that in case of turbulent
fluid motion this quantity represents the mean velocity.The dissipative
force which is nonlinear in the velocity has been expanded and developed up
to the third-order term. The resulting formula contains equilibrium time
correlation functions up to fourth order. Mulitlinear mode coupling theory
has been applied to calculate these quantities. Moreover, the formalism is
restricted to shear motion (no sound and heat flow effects). The result in
geometric space are formulas (\ref{10.9}) to (\ref{10.14}).

A major restriction of the investigation is that the N-particle theory in
its present form is with no boundaries. This is not just a theoretical
argument. The formulas have been applied to circular tube flow with the
result that the third order friction term is zero. This probably means that
the theory in this form is unable to describe turbulent tube flow. -
Presently, it is intended to test the theory with the empirical date of the
circular jet flow; since this is a configuration where the influence of the
boundary may be rather small. As a first attempt, the self-similar jet flow
formulas by Schlichting \cite{sch} have been used. A boundary layer
approximated form of the theoretical formulas has been explicitly calculated
for a point on the jet axis. When the Schlichting formulas are inserted,
some of the space integrals in (\ref{10.9})\ diverge in the region far from
the nozzle. Thus, the theory is in contradiction with these empirical
formulas.

An explanation can perhaps be found in the observation that at least for
laminar flow of low Reynolds number in the region far from the nozzle the
round jet transforms into a recirculation current. This has been
experimentally shown by Zauner \cite{z85}\ and theoretically calculated by
Schneider and co-workers \cite{sch85}, \cite{szb87}. Since, by this
configuration, the flow essentially extends over a finite space region, the
divergence mentioned above cannot occur. As a next step, therefore, it is
intended to investigate in which way the flow characteristics change when
the third order friction part is included into the calculation.

\appendix

\section{Appendix:\ Detailed formulas}

In this appendix, formulas are presented which appear in the course of the
analysis in the main part of the paper, but are too voluminous to be given
there. - First, the terms of the second order functional derivative of the
stress tensor kernel function S are written, as calculated from (\ref{3.1}),
(\ref{3.4}). The detailed formula for the second derivative reads:

\begin{equation}
\frac{\delta ^{2}S_{abcd}(\mathbf{x}^{\prime },t^{\prime })}{\delta b_{e}(%
\mathbf{x}^{\prime \prime },t^{\prime \prime })\delta b_{f}(\mathbf{x}%
^{\prime \prime \prime },t^{\prime \prime \prime })}  \label{a.1}
\end{equation}

For shortness, on the left hand side, indices and variables are omitted.
Second order derivatives with respect to the momentum part of the conjugated
parameters are needed only (that is, the greek index $\epsilon $\ is
switched to $e$). For each term, the first step is just the definition of
the quantity.

\begin{align}
& \left[ \frac{\delta ^{2}S}{\delta b\,\delta b}\right] ^{(p,1)}=-\delta
(t^{\prime \prime }-t^{\prime })\beta \,\func{tr}\{\frac{\delta
f_{L}(t^{\prime })}{\delta b_{f}(\mathbf{x}^{\prime \prime \prime
},t^{\prime \prime \prime })}[\mathcal{\tilde{G}}(t^{\prime },t)\mathcal{Q}%
(t)s_{ac}(\mathbf{x})][\mathcal{Q}(t^{\prime })s_{bd}(\mathbf{x}^{\prime
})]\delta a_{e}(\mathbf{x}^{\prime \prime },t^{\prime })\}  \notag \\
& =\delta (t^{\prime \prime \prime }-t^{\prime })\delta (t^{\prime \prime
}-t^{\prime })\beta \,\langle \lbrack \mathcal{\tilde{G}}(t^{\prime },t)%
\hat{s}_{ac}(\mathbf{x},t)][\mathcal{Q}(t^{\prime })[\mathcal{Q}(t^{\prime
})s_{bd}(\mathbf{x}^{\prime })]\delta a_{e}(\mathbf{x}^{\prime \prime
},t^{\prime })]\delta a_{f}(\mathbf{x}^{\prime \prime \prime },t^{\prime
})\rangle _{L,t^{\prime }}  \label{a.2}
\end{align}

\begin{eqnarray}
&&\left[ \frac{\delta ^{2}S}{\delta b\,\delta b}\right] ^{(p,2)}=-\delta
(t^{\prime \prime }-t^{\prime })\beta \,\langle \lbrack \mathcal{\tilde{G}}%
(t^{\prime },t)\mathcal{Q}(t)s_{ac}(\mathbf{x})]\mathcal{Q}(t^{\prime })[%
\mathcal{Q}(t^{\prime })s_{bd}(\mathbf{x}^{\prime })]\frac{\delta (\delta
a_{e}(\mathbf{x}^{\prime \prime },t^{\prime }))}{\delta b_{f}(\mathbf{x}%
^{\prime \prime \prime },t^{\prime \prime \prime })}\rangle _{L,t^{\prime }}
\notag \\
&=&-\delta (t^{\prime \prime \prime }-t^{\prime })\delta (t^{\prime \prime
}-t^{\prime })\beta \,\langle \lbrack \mathcal{\tilde{G}}(t^{\prime },t)%
\mathcal{Q}(t)s_{ac}(\mathbf{x})]\mathcal{Q}(t^{\prime })s_{bd}(\mathbf{x}%
^{\prime })\rangle _{L,t^{\prime }}\langle a_{e}(\mathbf{x}^{\prime \prime
})\delta a_{f}(\mathbf{x}^{\prime \prime \prime },t^{\prime })\rangle
_{L,t^{\prime }}  \label{a.3}
\end{eqnarray}%
\begin{align}
& \left[ \frac{\delta ^{2}S}{\delta b\,\delta b}\right] ^{(p,3)}=-\delta
(t^{\prime \prime }-t^{\prime })\beta \,\langle \lbrack \mathcal{\tilde{G}}%
(t^{\prime },t)\mathcal{Q}(t)s_{ac}(\mathbf{x})]\mathcal{Q}(t^{\prime })[%
\frac{\delta \mathcal{Q}(t^{\prime })}{\delta b_{f}(\mathbf{x}^{\prime
\prime \prime },t^{\prime \prime \prime })}s_{bd}(\mathbf{x}^{\prime
})]\delta a_{e}(\mathbf{x}^{\prime \prime },t^{\prime })\rangle
_{L,t^{\prime }}  \notag \\
& =-\delta (t^{\prime \prime \prime }-t^{\prime })\delta (t^{\prime \prime
}-t^{\prime })\times  \notag \\
& \text{ \ \ \ \ \ \ \ \ \ \ \ \ \ \ \ }\times \beta \langle \lbrack 
\mathcal{\tilde{G}}(t^{\prime },t)\mathcal{Q}(t)s_{ac}(\mathbf{x})]\mathcal{Q%
}(t^{\prime })[\mathcal{P}(t^{\prime })\delta a_{f}(\mathbf{x}^{\prime
\prime \prime },t^{\prime })\mathcal{Q}(t^{\prime })s_{bd}(\mathbf{x}%
^{\prime })]\delta a_{e}(\mathbf{x}^{\prime \prime },t^{\prime })\rangle
_{L,t^{\prime }}  \label{a.4}
\end{align}%
\begin{align}
& \left[ \frac{\delta ^{2}S}{\delta b\,\delta b}\right] ^{(p,4)}=-\delta
(t^{\prime \prime }-t^{\prime })\beta \,\langle \lbrack \mathcal{\tilde{G}}%
(t^{\prime },t)\mathcal{Q}(t)s_{ac}(\mathbf{x})]\frac{\delta \mathcal{Q}%
(t^{\prime })}{\delta b_{f}(\mathbf{x}^{\prime \prime \prime },t^{\prime
\prime \prime })}[\mathcal{Q}(t^{\prime })s_{bd}(\mathbf{x}^{\prime
})]\delta a_{e}(\mathbf{x}^{\prime \prime },t^{\prime })\rangle
_{L,t^{\prime }}  \notag \\
& -\delta (t^{\prime \prime \prime }-t^{\prime })\delta (t^{\prime \prime
}-t^{\prime })\times  \notag \\
& \text{ \ \ \ \ \ \ \ \ \ \ \ \ \ \ \ }\times \beta \,\langle \lbrack 
\mathcal{\tilde{G}}(t^{\prime },t)\mathcal{Q}(t)s_{ac}(\mathbf{x})]\mathcal{P%
}(t^{\prime })\delta a_{f}(\mathbf{x}^{\prime \prime \prime },t^{\prime })%
\mathcal{Q}(t^{\prime })[\mathcal{Q}(t^{\prime })s_{bd}(\mathbf{x}^{\prime
})]\delta a_{e}(\mathbf{x}^{\prime \prime },t^{\prime })\rangle
_{L,t^{\prime }}  \label{a.5}
\end{align}%
\begin{align}
\left[ \frac{\delta ^{2}S}{\delta b\,\delta b}\right] ^{(p,5)}& =-\delta
(t^{\prime \prime }-t^{\prime })\beta \,\langle \lbrack \mathcal{\tilde{G}}%
(t^{\prime },t)\frac{\delta \mathcal{Q}(t)}{\delta b_{f}(\mathbf{x}^{\prime
\prime \prime },t^{\prime \prime \prime })}s_{ac}(\mathbf{x})]\mathcal{Q}%
(t^{\prime })[\mathcal{Q}(t^{\prime })s_{bd}(\mathbf{x}^{\prime })]\delta
a_{e}(\mathbf{x}^{\prime \prime },t^{\prime })\rangle _{L,t^{\prime }} 
\notag \\
& =-\delta (t^{\prime \prime \prime }-t)\delta (t^{\prime \prime }-t^{\prime
})\times  \notag \\
& \beta \,\langle \lbrack \mathcal{\tilde{G}}(t^{\prime },t)\mathcal{P}%
(t)\delta a_{f}(\mathbf{x}^{\prime \prime \prime },t)\mathcal{Q}(t)s_{ac}(%
\mathbf{x})]\mathcal{Q}(t^{\prime })[\mathcal{Q}(t^{\prime })s_{bd}(\mathbf{x%
}^{\prime })]\delta a_{e}(\mathbf{x}^{\prime \prime },t^{\prime })\rangle
_{L,t^{\prime }}  \notag \\
& =0  \label{a.6}
\end{align}

The last step is correct because of (\ref{3.6}) and $\mathcal{QP}=0$.

\begin{align}
& \left[ \frac{\delta ^{2}S}{\delta b\,\delta b}\right] ^{(p,6)}=-\delta
(t^{\prime \prime }-t^{\prime })\beta \,\langle \lbrack \frac{\delta 
\mathcal{\tilde{G}}(t^{\prime },t)}{\delta b_{f}(\mathbf{x}^{\prime \prime
\prime },t^{\prime \prime \prime })}\mathcal{Q}(t)s_{ac}(\mathbf{x})]%
\mathcal{Q}(t^{\prime })[\mathcal{Q}(t^{\prime })s_{bd}(\mathbf{x}^{\prime
})]\delta a_{e}(\mathbf{x}^{\prime \prime },t^{\prime })\rangle
_{L,t^{\prime }}  \notag \\
& =-\delta (t^{\prime \prime }-t^{\prime })\Theta (t^{\prime \prime \prime
}-t^{\prime })\Theta (t-t^{\prime \prime \prime })\times  \notag \\
& \times \beta \,\langle \lbrack \mathcal{\tilde{G}}(t^{\prime },t^{\prime
\prime \prime })\mathcal{LP}(t^{\prime \prime \prime })\delta a_{f}(\mathbf{x%
}^{\prime \prime \prime },t^{\prime \prime \prime })\mathcal{Q}(t^{\prime
\prime \prime })\mathcal{\tilde{G}}(t^{\prime \prime \prime },t)\mathcal{Q}%
(t)s_{ac}(\mathbf{x})]\mathcal{Q}(t^{\prime })[\mathcal{Q}(t^{\prime
})s_{bd}(\mathbf{x}^{\prime })]\delta a_{e}(\mathbf{x}^{\prime \prime
},t^{\prime })\rangle _{L,t^{\prime }}  \label{a.8}
\end{align}%
\begin{eqnarray}
&&\left[ \frac{\delta ^{2}S}{\delta b\,\delta b}\right] ^{(f,1)}=\Theta
(t^{\prime \prime }-t^{\prime })\Theta (t-t^{\prime \prime })\times  \notag
\\
&&\times \beta \func{tr}\{\frac{\delta f_{L}(t^{\prime })}{\delta b_{f}(%
\mathbf{x}^{\prime \prime \prime },t^{\prime \prime \prime })}[\mathcal{%
\tilde{G}}(t^{\prime },t^{\prime \prime })\mathcal{LP}(t^{\prime \prime }%
\mathcal{)}\delta a_{e}(\mathbf{x}^{\prime \prime },t^{\prime \prime })%
\mathcal{Q}(t^{\prime \prime }\mathcal{)\tilde{G}}(t^{\prime \prime },t)%
\mathcal{Q}(t\mathcal{)}s_{ac}(\mathbf{x})]\mathcal{Q}(t^{\prime }\mathcal{)}%
s_{bd}(\mathbf{x}^{\prime })\}  \notag \\
&=&-\delta (t^{\prime \prime \prime }-t^{\prime })\Theta (t^{\prime \prime
}-t^{\prime })\Theta (t-t^{\prime \prime })\times  \notag \\
&&\times \beta \langle \lbrack \mathcal{\tilde{G}}(t^{\prime },t^{\prime
\prime })\mathcal{LP}(t^{\prime \prime }\mathcal{)}\delta a_{e}(\mathbf{x}%
^{\prime \prime },t^{\prime \prime })\mathcal{Q}(t^{\prime \prime }\mathcal{)%
\tilde{G}}(t^{\prime \prime },t)\mathcal{Q}(t\mathcal{)}s_{ac}(\mathbf{x})][%
\mathcal{Q}(t^{\prime }\mathcal{)}s_{bd}(\mathbf{x}^{\prime })]\delta a_{f}(%
\mathbf{x}^{\prime \prime \prime },t^{\prime })\rangle _{L,t^{\prime }}
\label{a.9}
\end{eqnarray}%
\begin{eqnarray}
&&\left[ \frac{\delta ^{2}S}{\delta b\,\delta b}\right] ^{(f,2)}=\Theta
(t^{\prime \prime }-t^{\prime })\Theta (t-t^{\prime \prime })\times  \notag
\\
&&\times \beta \langle \lbrack \mathcal{\tilde{G}}(t^{\prime },t^{\prime
\prime })\mathcal{LP}(t^{\prime \prime }\mathcal{)}\delta a_{e}(\mathbf{x}%
^{\prime \prime },t^{\prime \prime })\mathcal{Q}(t^{\prime \prime }\mathcal{)%
\tilde{G}}(t^{\prime \prime },t)\mathcal{Q}(t\mathcal{)}s_{ac}(\mathbf{x})]%
\frac{\delta \mathcal{Q}(t^{\prime })}{\delta b_{f}(\mathbf{x}^{\prime
\prime \prime },t^{\prime \prime \prime })}s_{bd}(\mathbf{x}^{\prime
})\rangle _{L,t^{\prime }}  \notag \\
&=&\delta (t^{\prime \prime \prime }-t^{\prime })\Theta (t^{\prime \prime
}-t^{\prime })\Theta (t-t^{\prime \prime })\times  \notag \\
&&\times \beta \langle \mathcal{\tilde{G}}[(t^{\prime },t^{\prime \prime })%
\mathcal{LP}(t^{\prime \prime }\mathcal{)}\delta a_{e}(\mathbf{x}^{\prime
\prime },t^{\prime \prime })\mathcal{Q}(t^{\prime \prime }\mathcal{)\tilde{G}%
}(t^{\prime \prime },t)\mathcal{Q}(t\mathcal{)}s_{ac}(\mathbf{x})]\mathcal{P}%
(t^{\prime })\delta a_{f}(\mathbf{x}^{\prime \prime \prime },t^{\prime })%
\mathcal{Q}(t^{\prime }\mathcal{)}s_{bd}(\mathbf{x}^{\prime })\rangle
_{L,t^{\prime }}  \label{a.10}
\end{eqnarray}%
\begin{eqnarray}
&&\left[ \frac{\delta ^{2}S}{\delta b\,\delta b}\right] ^{(f,3)}=\Theta
(t^{\prime \prime }-t^{\prime })\Theta (t-t^{\prime \prime })\times  \notag
\\
&&\times \beta \langle \lbrack \mathcal{\tilde{G}}(t^{\prime },t^{\prime
\prime })\mathcal{LP}(t^{\prime \prime }\mathcal{)}\delta a_{e}(\mathbf{x}%
^{\prime \prime },t^{\prime \prime })\mathcal{Q}(t^{\prime \prime }\mathcal{)%
\tilde{G}}(t^{\prime \prime },t)\frac{\delta \mathcal{Q}(t)}{\delta b_{f}(%
\mathbf{x}^{\prime \prime \prime },t^{\prime \prime \prime })}s_{ac}(\mathbf{%
x})]\mathcal{Q}(t^{\prime }\mathcal{)}s_{bd}(\mathbf{x}^{\prime })\rangle
_{L,t^{\prime }}  \notag \\
&=&\delta (t^{\prime \prime \prime }-t)\Theta (t^{\prime \prime }-t^{\prime
})\Theta (t-t^{\prime \prime })\times  \notag \\
&&\times \beta \langle \lbrack \mathcal{\tilde{G}}(t^{\prime },t^{\prime
\prime })\mathcal{LP}(t^{\prime \prime }\mathcal{)}\delta a_{e}(\mathbf{x}%
^{\prime \prime },t^{\prime \prime })\mathcal{Q}(t^{\prime \prime }\mathcal{)%
\tilde{G}}(t^{\prime \prime },t)\mathcal{P}(t)\delta a_{f}(\mathbf{x}%
^{\prime \prime \prime },t)\mathcal{Q}(t\mathcal{)}s_{ac}(\mathbf{x})]%
\mathcal{Q}(t^{\prime }\mathcal{)}s_{bd}(\mathbf{x}^{\prime })\rangle
_{L,t^{\prime }}  \notag \\
&=&0  \label{a.11}
\end{eqnarray}%
\begin{eqnarray}
&&\left[ \frac{\delta ^{2}S}{\delta b\,\delta b}\right] ^{(f,4)}=\Theta
(t^{\prime \prime }-t^{\prime })\Theta (t-t^{\prime \prime })\times  \notag
\\
&&\times \beta \langle \lbrack \mathcal{\tilde{G}}(t^{\prime },t^{\prime
\prime })\mathcal{LP}(t^{\prime \prime }\mathcal{)}\delta a_{e}(\mathbf{x}%
^{\prime \prime },t^{\prime \prime })\frac{\delta \mathcal{Q}(t^{\prime
\prime })}{\delta b_{f}(\mathbf{x}^{\prime \prime \prime },t^{\prime \prime
\prime })}\mathcal{\tilde{G}}(t^{\prime \prime },t)\mathcal{Q}(t\mathcal{)}%
s_{ac}(\mathbf{x})]\mathcal{Q}(t^{\prime }\mathcal{)}s_{bd}(\mathbf{x}%
^{\prime })\rangle _{L,t^{\prime }}  \notag \\
&=&\delta (t^{\prime \prime \prime }-t^{\prime \prime })\Theta (t^{\prime
\prime }-t^{\prime })\Theta (t-t^{\prime \prime })\times  \notag \\
&&\times \beta \langle \mathcal{\tilde{G}}[(t^{\prime },t^{\prime \prime })%
\mathcal{LP}(t^{\prime \prime }\mathcal{)}\delta a_{e}(\mathbf{x}^{\prime
\prime },t^{\prime \prime })\mathcal{P}(t^{\prime \prime })\delta a_{f}(%
\mathbf{x}^{\prime \prime \prime },t^{\prime \prime })\mathcal{Q}(t^{\prime
\prime })\mathcal{\tilde{G}}(t^{\prime \prime },t)\mathcal{Q}(t\mathcal{)}%
s_{ac}(\mathbf{x})]\mathcal{Q}(t^{\prime }\mathcal{)}s_{bd}(\mathbf{x}%
^{\prime })\rangle _{L,t^{\prime }}  \label{a.13}
\end{eqnarray}%
\begin{eqnarray}
&&\left[ \frac{\delta ^{2}S}{\delta b\,\delta b}\right] ^{(f,5)}=\Theta
(t^{\prime \prime }-t^{\prime })\Theta (t-t^{\prime \prime })\times  \notag
\\
&&\times \beta \langle \lbrack \mathcal{\tilde{G}}(t^{\prime },t^{\prime
\prime })\mathcal{LP}(t^{\prime \prime }\mathcal{)}\frac{\delta (\delta
a_{e}(\mathbf{x}^{\prime \prime },t^{\prime \prime }))}{\delta b_{f}(\mathbf{%
x}^{\prime \prime \prime },t^{\prime \prime \prime })}\mathcal{Q}(t^{\prime
\prime }\mathcal{)\tilde{G}}(t^{\prime \prime },t)\mathcal{Q}(t\mathcal{)}%
s_{ac}(\mathbf{x})]\mathcal{Q}(t^{\prime }\mathcal{)}s_{bd}(\mathbf{x}%
^{\prime })\rangle _{L,t^{\prime }}  \notag \\
&=&-\Theta (t^{\prime \prime }-t^{\prime })\Theta (t-t^{\prime \prime
})\times  \notag \\
&&\times \beta \langle \lbrack \mathcal{\tilde{G}}(t^{\prime },t^{\prime
\prime })\mathcal{LP}(t^{\prime \prime }\mathcal{)}\frac{\delta \langle
a_{e}(x^{\prime \prime })\rangle _{L,t^{\prime \prime }}}{\delta b_{f}(%
\mathbf{x}^{\prime \prime \prime },t^{\prime \prime \prime })}\mathcal{Q}%
(t^{\prime \prime }\mathcal{)\tilde{G}}(t^{\prime \prime },t)\mathcal{Q}(t%
\mathcal{)}s_{ac}(\mathbf{x})]\mathcal{Q}(t^{\prime }\mathcal{)}s_{bd}(%
\mathbf{x}^{\prime })\rangle _{L,t^{\prime }}  \notag \\
&=&\delta (t^{\prime \prime \prime }-t^{\prime \prime })\Theta (t^{\prime
\prime }-t^{\prime })\Theta (t-t^{\prime \prime })\times  \notag \\
&&\times \beta \langle \lbrack \mathcal{\tilde{G}}(t^{\prime },t^{\prime
\prime })\mathcal{LP}(t^{\prime \prime }\mathcal{)Q}(t^{\prime \prime }%
\mathcal{)\tilde{G}}(t^{\prime \prime },t)\mathcal{Q}(t\mathcal{)}s_{ac}(%
\mathbf{x})]\mathcal{Q}(t^{\prime }\mathcal{)}s_{bd}(\mathbf{x}^{\prime
})\rangle _{L,t^{\prime }}\langle a_{e}(x^{\prime \prime })\delta
a_{f}(x^{\prime \prime \prime },t^{\prime \prime })\rangle _{L,t^{\prime
\prime }}  \notag \\
&=&0  \label{a.14}
\end{eqnarray}

\begin{eqnarray}
&&\left[ \frac{\delta ^{2}S}{\delta b\,\delta b}\right] ^{(f,6)}=\Theta
(t^{\prime \prime }-t^{\prime })\Theta (t-t^{\prime \prime })\times  \notag
\\
&&\times \beta \langle \lbrack \mathcal{\tilde{G}}(t^{\prime },t^{\prime
\prime })\mathcal{L}\frac{\delta \mathcal{P}(t^{\prime \prime })}{\delta
b_{f}(\mathbf{x}^{\prime \prime \prime },t^{\prime \prime \prime })}\delta
a_{e}(x^{\prime \prime },t^{\prime \prime })\mathcal{Q}(t^{\prime \prime }%
\mathcal{)\tilde{G}}(t^{\prime \prime },t)\mathcal{Q}(t\mathcal{)}\hat{s}%
_{ac}(\mathbf{x})]\mathcal{Q}(t^{\prime }\mathcal{)}\hat{s}_{bd}(\mathbf{x}%
^{\prime })\rangle _{L,t^{\prime }}  \notag \\
&=&-\delta (t^{\prime \prime \prime }-t^{\prime \prime })\Theta (t^{\prime
\prime }-t^{\prime })\Theta (t-t^{\prime \prime })\times  \notag \\
&&\times \beta \langle \lbrack \mathcal{\tilde{G}}(t^{\prime },t^{\prime
\prime })\mathcal{LP}(t^{\prime \prime }\mathcal{)}\delta a_{f}(\mathbf{x}%
^{\prime \prime \prime },t^{\prime \prime })\mathcal{Q}(t^{\prime \prime }%
\mathcal{)}\delta a_{e}(x^{\prime \prime },t^{\prime \prime })\mathcal{Q}%
(t^{\prime \prime }\mathcal{)\tilde{G}}(t^{\prime \prime },t)\mathcal{Q}(t%
\mathcal{)}\hat{s}_{ac}(\mathbf{x})]\times  \notag \\
&&\text{ \ \ \ \ \ \ \ \ \ \ \ \ \ \ \ \ \ \ \ \ \ \ \ \ \ \ \ \ \ \ \ \ \ \
\ \ \ \ \ \ \ \ \ \ \ \ \ \ \ \ \ \ \ \ \ \ \ \ \ \ \ \ \ \ \ \ \ \ \ \ \ \ }%
\times \mathcal{Q}(t^{\prime }\mathcal{)}\hat{s}_{bd}(\mathbf{x}^{\prime
})\rangle _{L,t^{\prime }}  \label{a.15}
\end{eqnarray}%
\begin{eqnarray}
&&\left[ \frac{\delta ^{2}S}{\delta b\,\delta b}\right] ^{(f,7)}=\Theta
(t^{\prime \prime }-t^{\prime })\Theta (t-t^{\prime \prime })\times  \notag
\\
&&\times \beta \langle \lbrack \mathcal{\tilde{G}}(t^{\prime },t^{\prime
\prime })\mathcal{LP}(t^{\prime \prime }\mathcal{)}\delta a_{e}(x^{\prime
\prime },t^{\prime \prime })\mathcal{Q}(t^{\prime \prime }\mathcal{)}\frac{%
\delta \mathcal{\tilde{G}}(t^{\prime \prime },t)}{\delta b_{f}(\mathbf{x}%
^{\prime \prime \prime },t^{\prime \prime \prime })}\mathcal{Q}(t\mathcal{)}%
s_{ac}(\mathbf{x})]\mathcal{Q}(t^{\prime }\mathcal{)}s_{bd}(\mathbf{x}%
^{\prime })\rangle _{L,t^{\prime }}  \notag \\
&=&\Theta (t^{\prime \prime \prime }-t^{\prime \prime })\Theta (t-t^{\prime
\prime \prime })\Theta (t^{\prime \prime }-t^{\prime })\Theta (t-t^{\prime
\prime })\times  \notag \\
&&\times \beta \langle \lbrack \mathcal{\tilde{G}}(t^{\prime },t^{\prime
\prime })\mathcal{LP}(t^{\prime \prime }\mathcal{)}\delta a_{e}(x^{\prime
\prime },t^{\prime \prime })\mathcal{Q}(t^{\prime \prime }\mathcal{)\times }
\notag \\
&&\text{ \ \ \ \ \ \ \ \ \ \ \ \ \ \ }\mathcal{\times \tilde{G}}(t^{\prime
\prime },t^{\prime \prime \prime })\mathcal{LP}(t^{\prime \prime \prime
})\delta a_{f}(x^{\prime \prime \prime },t^{\prime \prime \prime })\mathcal{Q%
}(t^{\prime \prime \prime }))\mathcal{\tilde{G}}(t^{\prime \prime \prime },t)%
\mathcal{Q}(t\mathcal{)}s_{ac}(\mathbf{x})]\mathcal{Q}(t^{\prime }\mathcal{)}%
s_{bd}(\mathbf{x}^{\prime })\rangle _{L,t^{\prime }}  \label{a.16}
\end{eqnarray}%
\begin{eqnarray}
&&\left[ \frac{\delta ^{2}S}{\delta b\,\delta b}\right] ^{(f,8)}=\Theta
(t^{\prime \prime }-t^{\prime })\Theta (t-t^{\prime \prime })\times  \notag
\\
&&\times \beta \langle \lbrack \frac{\delta \mathcal{\tilde{G}}(t^{\prime
},t^{\prime \prime })}{\delta b_{f}(\mathbf{x}^{\prime \prime \prime
},t^{\prime \prime \prime })}\mathcal{LP}(t^{\prime \prime }\mathcal{)}%
\delta a_{e}(x^{\prime \prime },t^{\prime \prime })\mathcal{Q}(t^{\prime
\prime }\mathcal{)\tilde{G}}(t^{\prime \prime },t)\mathcal{Q}(t\mathcal{)}%
s_{ac}(\mathbf{x})]\mathcal{Q}(t^{\prime }\mathcal{)}s_{bd}(\mathbf{x}%
^{\prime })\rangle _{L,t^{\prime }}  \notag \\
&=&\Theta (t^{\prime \prime \prime }-t^{\prime })\Theta (t^{\prime \prime
}-t^{\prime \prime \prime })\Theta (t^{\prime \prime }-t^{\prime })\Theta
(t-t^{\prime \prime })\times  \notag \\
&&\times \beta \langle \lbrack \mathcal{\tilde{G}}(t^{\prime },t^{\prime
\prime \prime })\mathcal{LP}(t^{\prime \prime \prime }\mathcal{)}\delta
a_{f}(\mathbf{x}^{\prime \prime \prime },t^{\prime \prime \prime })\mathcal{Q%
}(t^{\prime \prime \prime }\mathcal{)})\mathcal{\tilde{G}}(t^{\prime \prime
\prime },t^{\prime \prime })\times  \notag \\
&&\text{ \ \ \ \ \ \ \ \ \ \ \ \ \ \ \ \ \ \ \ \ \ \ \ }\times \mathcal{LP}%
(t^{\prime \prime }\mathcal{)}\delta a_{e}(x^{\prime \prime },t^{\prime
\prime })\mathcal{Q}(t^{\prime \prime }\mathcal{)\tilde{G}}(t^{\prime \prime
},t)\mathcal{Q}(t\mathcal{)}s_{ac}(\mathbf{x})]\mathcal{Q}(t^{\prime }%
\mathcal{)}s_{bd}(\mathbf{x}^{\prime })\rangle _{L,t^{\prime }}  \label{a.17}
\end{eqnarray}

Results (\ref{a.11}), (\ref{a.14}) are found the same way as (\ref{a.6}). -
Below, the non-zero terms are presented in the form they assume after taking 
$b=b_{0}$, and reformulation, in sec. 3:

\begin{equation}
\left[ \frac{\delta ^{2}R}{\delta b\,\delta b}\right] _{0}^{(p,1)}=-\delta
(t^{\prime \prime \prime }-t^{\prime })\delta (t^{\prime \prime }-t^{\prime
})\langle \lbrack \mathcal{P}r_{0}+\mathcal{L}g(t-t^{\prime })h_{0}][%
\mathcal{Q}\hat{w}_{1}^{\ast }h_{2}^{\ast }]h_{3}^{\ast }\rangle _{0}\,
\label{a.32}
\end{equation}%
\begin{equation}
\left[ \frac{\delta ^{2}R}{\delta b\,\delta b}\right] _{0}^{(p,2)}=-\delta
(t^{\prime \prime \prime }-t^{\prime })\delta (t^{\prime \prime }-t^{\prime
})\,\langle \lbrack g(t-t^{\prime })\hat{r}_{0}]\hat{r}_{1}^{\ast }\rangle
_{0}\langle h_{2}^{\ast }h_{3}^{\ast }\rangle _{0}  \label{a.33}
\end{equation}

\begin{equation}
\left[ \frac{\delta ^{2}R}{\delta b\,\delta b}\right] _{0}^{(p,3)}=-\delta
(t^{\prime \prime \prime }-t^{\prime })\delta (t^{\prime \prime }-t^{\prime
})\,\langle \lbrack g(t-t^{\prime })\hat{r}_{0}][\mathcal{P}\hat{r}%
_{1}^{\ast }h_{3}^{\ast }]h_{2}^{\ast }\rangle _{0}  \label{a.34}
\end{equation}

\begin{equation}
\left[ \frac{\delta ^{2}R}{\delta b\,\delta b}\right] _{0}^{(p,4)}=\delta
(t^{\prime \prime \prime }-t^{\prime })\delta (t^{\prime \prime }-t^{\prime
})\,\langle \lbrack \mathcal{P}r_{0}+\mathcal{L}g(t-t^{\prime })h_{0}]%
\mathcal{P}h_{3}^{\ast }\mathcal{Q}\hat{r}_{1}^{\ast }h_{2}^{\ast }\rangle
_{0}  \label{a.35}
\end{equation}%
\begin{equation}
\left[ \frac{\delta ^{2}R}{\delta b\,\delta b}\right] _{0}^{(p,6)}=-\delta
(t^{\prime \prime }-t^{\prime })\Theta (t^{\prime \prime \prime }-t^{\prime
})\Theta (t-t^{\prime \prime \prime })\langle g[(t^{\prime \prime \prime
}-t^{\prime })\mathcal{QLP}h_{3}^{\ast }g(t-t^{\prime \prime \prime })\hat{r}%
_{0}]\hat{r}_{1}^{\ast }h_{2}^{\ast }\rangle _{0}  \label{a.36}
\end{equation}

\begin{equation}
\left[ \frac{\delta ^{2}R}{\delta b\,\delta b}\right] _{0}^{(f,1)}=-\delta
(t^{\prime \prime \prime }-t^{\prime })\Theta (t^{\prime \prime }-t^{\prime
})\Theta (t-t^{\prime \prime })\langle \lbrack \mathcal{L}g(t^{\prime \prime
}-t^{\prime })\mathcal{P}h_{2}^{\ast }g(t-t^{\prime \prime })\hat{r}_{0}]%
\hat{r}_{1}^{\ast }h_{3}^{\ast }\rangle _{0}  \label{a.37}
\end{equation}%
\begin{equation}
\left[ \frac{\delta ^{2}R}{\delta b\,\delta b}\right] _{0}^{(f,2)}=\delta
(t^{\prime \prime \prime }-t^{\prime })\Theta (t^{\prime \prime }-t^{\prime
})\Theta (t-t^{\prime \prime })\langle \lbrack \mathcal{L}g(t^{\prime \prime
}-t^{\prime })\mathcal{P}h_{2}^{\ast }g(t-t^{\prime \prime })\hat{r}_{0}]%
\mathcal{P}\hat{r}_{1}^{\ast }h_{3}^{\ast }\rangle _{0}  \label{a.38}
\end{equation}%
\begin{equation}
\left[ \frac{\delta ^{2}R}{\delta b\,\delta b}\right] _{0}^{(f,4)}=\delta
(t^{\prime \prime \prime }-t^{\prime \prime })\Theta (t^{\prime \prime
}-t^{\prime })\Theta (t-t^{\prime \prime })\langle \lbrack g(t^{\prime
\prime }-t^{\prime })\mathcal{QLP}h_{2}^{\ast }\mathcal{P}h_{3}^{\ast
}g(t-t^{\prime \prime })\hat{r}_{0}]\hat{r}_{1}^{\ast }\rangle _{0}
\label{a.39}
\end{equation}

\begin{equation}
\left[ \frac{\delta ^{2}R}{\delta b\,\delta b}\right] _{0}^{(f,6)}=-\delta
(t^{\prime \prime \prime }-t^{\prime \prime })\Theta (t^{\prime \prime
}-t^{\prime })\Theta (t-t^{\prime \prime })\langle \lbrack g(t^{\prime
\prime }-t^{\prime })\mathcal{QLP}h_{3}^{\ast }\mathcal{Q}h_{2}^{\ast
}g(t-t^{\prime \prime })\hat{r}_{0}]\hat{r}_{1}^{\ast }\rangle _{0}
\label{a.40}
\end{equation}%
\begin{eqnarray}
&&\left[ \frac{\delta ^{2}R}{\delta b\,\delta b}\right] _{0}^{(f,7)}=\Theta
(t^{\prime \prime \prime }-t^{\prime \prime })\Theta (t-t^{\prime \prime
\prime })\Theta (t^{\prime \prime }-t^{\prime })\Theta (t-t^{\prime \prime
})\times  \notag \\
&&\text{ \ \ \ \ \ \ \ \ \ \ \ \ \ \ \ \ \ \ \ \ \ \ \ \ \ \ }\times \langle
\lbrack g(t^{\prime \prime }-t^{\prime })\mathcal{QLP}h_{2}^{\ast
}g(t^{\prime \prime \prime }-t^{\prime \prime })\mathcal{QLP}h_{3}^{\ast
}g(t-t^{\prime \prime \prime })\hat{r}_{0}]\hat{r}_{1}^{\ast }\rangle _{0}
\label{a.41}
\end{eqnarray}%
\begin{eqnarray}
&&\left[ \frac{\delta ^{2}R}{\delta b\,\delta b}\right] _{0}^{(f,8)}=\Theta
(t^{\prime \prime \prime }-t^{\prime })\Theta (t^{\prime \prime }-t^{\prime
\prime \prime })\Theta (t^{\prime \prime }-t^{\prime })\Theta (t-t^{\prime
\prime })\times  \notag \\
&&\text{ \ \ \ \ \ \ \ \ \ \ \ \ \ \ \ \ \ \ \ \ \ \ \ \ \ \ \ }\times
\langle \lbrack g(t^{\prime \prime \prime }-t^{\prime })\mathcal{QLP}%
h_{3}^{\ast }g(t^{\prime \prime }-t^{\prime \prime \prime })\mathcal{QLP}%
h_{2}^{\ast }g(t-t^{\prime \prime })\hat{r}_{0}]\hat{r}_{1}^{\ast }\rangle
_{0}  \label{a.42}
\end{eqnarray}

After evaluation of the projection operators in section 4, we obtain for the
11 terms, now called $K_{1},\cdots ,K_{11}$ (Symbol $_{0}$ omitted from $%
\langle \,\rangle $):%
\begin{eqnarray}
(K_{1})_{0} &=&\langle \lbrack \mathcal{L}g(t-t_{1})h_{0}]h_{3}^{\ast
}h_{4}^{\ast }\rangle \langle \hat{r}_{1}^{\ast }h_{2}^{\ast }h_{4}\rangle
-\langle \lbrack \mathcal{L}g(t-t_{1})h_{0}]\hat{r}_{1}^{\ast }h_{2}^{\ast
}h_{3}^{\ast }\rangle  \notag \\
(K_{2})_{0} &=&-\langle \lbrack g(t-t_{1})\hat{r}_{0}]\hat{r}_{1}^{\ast
}\rangle \langle h_{2}^{\ast }h_{3}^{\ast }\rangle  \notag \\
(K_{3})_{0} &=&-\langle \lbrack g(t-t_{1})\hat{r}_{0}]h_{2}^{\ast
}h_{4}^{\ast }\rangle \langle \hat{r}_{1}^{\ast }h_{3}^{\ast }h_{4}\rangle 
\notag \\
(K_{4})_{0} &=&\langle \lbrack \mathcal{L}g(t-t_{1})h_{0}]h_{4}^{\ast
}\rangle \langle \hat{r}_{1}^{\ast }h_{2}^{\ast }h_{3}^{\ast }h_{4}\rangle
-\langle \lbrack \mathcal{L}g(t-t_{1})h_{0}]h_{4}^{\ast }\rangle \langle 
\hat{r}_{1}^{\ast }h_{2}^{\ast }h_{5}\rangle \langle h_{3}^{\ast
}h_{4}h_{5}^{\ast }\rangle  \notag \\
(K_{5})_{0} &=&\langle \lbrack g(t-t_{3})\hat{r}_{0}]h_{3}^{\ast
}h_{5}^{\ast }\rangle \langle \lbrack g(t_{3}-t_{1})\hat{r}_{5}]\hat{r}%
_{1}^{\ast }h_{2}^{\ast }\rangle  \notag \\
(K_{6})_{0} &=&-\langle \lbrack g(t-t_{2})\hat{r}_{0}]h_{2}^{\ast
}h_{4}^{\ast }\rangle \langle \lbrack \mathcal{L}g(t_{2}-t_{1})h_{4}]\hat{r}%
_{1}^{\ast }h_{3}^{\ast }\rangle  \notag \\
(K_{7})_{0} &=&\langle \lbrack g(t-t_{2})\hat{r}_{0}]h_{2}^{\ast
}h_{5}^{\ast }\rangle \langle \lbrack \mathcal{L}g(t_{2}-t_{1})h_{5}]h_{4}^{%
\ast }\rangle \langle \hat{r}_{1}^{\ast }h_{3}^{\ast }h_{4}\rangle
\label{a.43} \\
(K_{8})_{0} &=&-\langle \lbrack g(t-t_{2})\hat{r}_{0}]h_{3}^{\ast }\rangle
\langle \lbrack g(t_{2}-t_{1})\hat{r}_{5}]\hat{r}_{1}^{\ast }\rangle \langle
h_{2}^{\ast }h_{5}^{\ast }\rangle  \notag \\
&&+\langle \lbrack g(t-t_{2})\hat{r}_{0}]h_{3}^{\ast }h_{6}^{\ast }\rangle
\langle \lbrack g(t_{2}-t_{1})\hat{r}_{5}]\hat{r}_{1}^{\ast }\rangle \langle
h_{2}^{\ast }h_{5}^{\ast }h_{6}\rangle  \notag \\
(K_{9})_{0} &=&-\langle \lbrack \mathcal{L}g(t-t_{2})h_{0}]h_{2}^{\ast
}\rangle \langle \lbrack g(t_{2}-t_{1})\hat{r}_{5}]\hat{r}_{1}^{\ast
}\rangle \langle h_{3}^{\ast }h_{5}^{\ast }\rangle  \notag \\
&&+\langle \lbrack g(t-t_{2})\hat{r}_{0}]h_{2}^{\ast }h_{3}^{\ast
}h_{5}^{\ast }\rangle \langle \lbrack g(t_{2}-t_{1})\hat{r}_{5}]\hat{r}%
_{1}^{\ast }\rangle  \notag \\
&&+\langle \lbrack g(t-t_{2})\hat{r}_{0}]h_{2}^{\ast }h_{6}^{\ast }\rangle
\langle \lbrack g(t_{2}-t_{1})\hat{r}_{5}]\hat{r}_{1}^{\ast }\rangle \langle
h_{3}^{\ast }h_{5}^{\ast }h_{6}\rangle  \notag \\
(K_{10})_{0} &=&\langle \lbrack g(t-t_{3})\hat{r}_{0}]h_{3}^{\ast
}h_{7}^{\ast }\rangle \langle \lbrack g(t_{2}-t_{1})\hat{r}_{5}]\hat{r}%
_{1}^{\ast }\rangle \langle \lbrack g(t_{3}-t_{2})\hat{r}_{7}]h_{2}^{\ast
}h_{5}^{\ast }\rangle  \notag \\
(K_{11})_{0} &=&\langle \lbrack g(t-t_{2})\hat{r}_{0}]h_{2}^{\ast
}h_{7}^{\ast }\rangle \langle \lbrack g(t_{3}-t_{1})\hat{r}_{5}]\hat{r}%
_{1}^{\ast }\rangle \langle \lbrack g(t_{2}-t_{3})\hat{r}_{7}]h_{3}^{\ast
}h_{5}^{\ast }\rangle  \notag
\end{eqnarray}

In sec. 4, these terms are combined into 5 main parts:

\begin{eqnarray}
(M_{1})_{0123} &=&-(j_{2})_{23}\gamma _{01}\delta (t-t^{\prime })  \notag \\
&&-2(s_{3})_{124}(\kappa _{05}(C_{3})_{534}(t-t^{\prime })+\partial
_{t}(C_{3})_{034}(t-t^{\prime }))  \notag \\
&&-\omega _{18}((C_{4})_{7823}(t-t^{\prime })\kappa _{07}+\partial
_{t}(C_{4})_{0823}(t-t^{\prime }))  \notag \\
&&-\frac{1}{3}(\kappa _{07}\partial _{t}(C_{4})_{7123}(t-t^{\prime
})+\partial _{tt}(C_{4})_{0123}(t-t^{\prime }))  \notag \\
(M_{2})_{0123} &=&((C_{3})_{824}(t-t^{\prime \prime })\kappa _{08}+\partial
_{t}(C_{3})_{024}(t-t^{\prime \prime }))  \notag \\
&&\times (\kappa _{45}((s_{3})_{513}-(s_{3})_{135})  \notag \\
&&+2\omega _{16}((C_{3})_{763}(t^{\prime \prime }-t^{\prime })\kappa
_{47}+\partial _{t}(C_{3})_{463}(t^{\prime \prime }-t^{\prime })  \notag \\
&&+\kappa _{47}\partial _{t}(C_{3})_{713}(t^{\prime \prime }-t^{\prime })+%
\frac{1}{2}\partial _{tt}(C_{3})_{413}(t^{\prime \prime }-t^{\prime }))
\label{a.44} \\
(M_{3})_{0123} &=&\gamma _{51}(\kappa _{08}((C_{3})_{836}(t-t^{\prime \prime
})(j_{3})_{625}+(C_{3})_{826}(t-t^{\prime \prime })(j_{3})_{635})  \notag \\
&&+\partial _{t}(C_{3})_{026}(t-t^{\prime \prime })(j_{3})_{635}+\partial
_{t}(C_{3})_{036}(t-t^{\prime \prime })(j_{3})_{625}  \notag \\
&&-(C_{4})_{8235}(t-t^{\prime \prime })\kappa _{08}-\partial
_{t}(C_{4})_{0235}(t-t^{\prime \prime }))  \notag \\
(M_{4})_{0123} &=&\gamma _{51}(\kappa _{08}(C_{3})_{837}(t-t^{\prime \prime
\prime })+\partial _{t}(C_{3})_{037}(t-t^{\prime \prime \prime }))  \notag \\
&&\times (\kappa _{79}(C_{3})_{925}(t^{\prime \prime \prime }-t^{\prime
\prime })+\partial _{t}(C_{3})_{725}(t^{\prime \prime \prime }-t^{\prime
\prime })  \notag \\
(M_{5})_{0123} &=&\gamma _{51}(\kappa _{08}(C_{3})_{827}(t-t^{\prime \prime
})+\partial _{t}(C_{3})_{027}(t-t^{\prime \prime }))  \notag \\
&&\times (\kappa _{79}(C_{3})_{935}(t^{\prime \prime }-t^{\prime \prime
\prime })+\partial _{t}(C_{3})_{735}(t^{\prime \prime }-t^{\prime \prime
\prime })  \notag
\end{eqnarray}

When the $M$-quantities are inserted into (\ref{3.10}), some of the time
integrations can be performed; finally, two parts of the kernel function
remain:

\begin{eqnarray}
\Phi 1_{0123}(t^{\prime }) &=&-(j_{2})_{23}\gamma _{01}\delta (t^{\prime }) 
\notag \\
&&-2(s_{3})_{124}(\kappa _{05}(C_{3})_{534}(t^{\prime })+\partial
_{t}(C_{3})_{034}(t^{\prime }))  \notag \\
&&+2(j_{3})_{625}\gamma _{51}((\kappa _{08}((C_{3})_{836}(t^{\prime
})+\partial _{t}(C_{3})_{036}(t^{\prime }))  \notag \\
&&-\kappa _{51}((C_{4})_{8235}(t^{\prime })\kappa _{08}+\partial
_{t}(C_{4})_{0235}(t^{\prime }))  \notag \\
&&-\frac{1}{3}(\kappa _{07}\partial _{t}(C_{4})_{7123}(t^{\prime })+\partial
_{tt}(C_{4})_{0123}(t^{\prime }))  \notag \\
\Phi 2_{0123}(t^{\prime },t^{\prime \prime }) &=&2\gamma _{51}(\kappa
_{08}(C_{3})_{837}(t^{\prime })+\partial _{t}(C_{3})_{037}(t^{\prime }))
\label{a.45} \\
&&\times (\kappa _{79}(C_{3})_{925}(t^{\prime \prime })+\partial
_{t}(C_{3})_{725}(t^{\prime \prime }))  \notag \\
&&+((C_{3})_{824}(t^{\prime })\kappa _{08}+\partial
_{t}(C_{3})_{024}(t^{\prime }))  \notag \\
&&\times (\kappa _{45}((s_{3})_{513}-(s_{3})_{135})  \notag \\
&&+2\omega _{16}((C_{3})_{763}(t^{\prime \prime })\kappa _{47}+\partial
_{t}(C_{3})_{463}(t^{\prime \prime }))  \notag \\
&&+\kappa _{47}\partial _{t}(C_{3})_{713}(t^{\prime \prime })+\frac{1}{2}%
\partial _{tt}(C_{3})_{413}(t^{\prime \prime }))  \notag
\end{eqnarray}

In sec. 5, the part of the second-order (in $b-b_{0}$) term which adds to
the third-order velocity term is calculated. The kernel function reads:

\begin{eqnarray}
(K_{2})_{012}\left( t^{\prime }\right) &=&\gamma _{01}\delta (t^{\prime
})\langle h_{2}\rangle  \notag \\
&&+\gamma _{31}(\partial _{t^{\prime }}(C_{3})_{032}(t^{\prime })+\kappa
_{04}(C_{3})_{432}(t^{\prime }))  \notag \\
&&+\partial _{t^{\prime }t^{\prime }}(C_{3})_{012}(t^{\prime
})+i\,k_{c}^{\prime \prime }\partial _{t^{\prime
}}(C_{3})_{012}|_{"2"=c}(t^{\prime })+\omega _{14}\partial _{t^{\prime
}}(C_{3})_{042}(t^{\prime })  \notag \\
&&+\kappa _{03}(\partial _{t^{\prime }}(C_{3})_{312}(t^{\prime
})+i\,k_{c}^{\prime \prime }(C_{3})_{312}|_{"2"=c}(t^{\prime })+\omega
_{14}(C_{3})_{342}(t^{\prime }))  \label{a.46}
\end{eqnarray}

The index $_{"2"=c}$\ means that the component index of the number index 2
runs over the three values $c=3,4,5$; in contrast to the general
prescription of this section where it has the fixed value 1. - In the
explicit formulation in sec. 7, we obtain for the expression in (\ref{7.2}),(%
\ref{7.3}):%
\begin{align}
\Phi 1& _{0123}(\mathbf{\mathbf{k}^{\prime },\mathbf{k}^{\prime \prime },k}%
^{\prime \prime \prime },t^{\prime })=-(2\pi )^{3}\delta _{23}\delta (%
\mathbf{k-\mathbf{k}^{\prime }})\gamma _{01}(\mathbf{k})\delta (t^{\prime })
\notag \\
& -2s_{124}(\mathbf{\mathbf{k}^{\prime }})(\kappa _{05}(\mathbf{k}%
)(C_{3})_{534}(\mathbf{\mathbf{k},\mathbf{k}^{\prime \prime \prime },\mathbf{%
k}^{\prime }+k}^{\prime \prime },t^{\prime })+\partial _{t}(C_{3})_{034}((%
\mathbf{\mathbf{k},\mathbf{k}^{\prime \prime \prime },\mathbf{k}^{\prime }+k}%
^{\prime \prime },t^{\prime }))  \notag \\
& +2j_{625}\gamma _{51}(\mathbf{\mathbf{k}^{\prime }})(\kappa _{05}(\mathbf{%
\mathbf{k}})(C_{3})_{536}(\mathbf{\mathbf{k},\mathbf{k}^{\prime \prime
\prime },\mathbf{k}^{\prime }+k}^{\prime \prime },t^{\prime })+\partial
_{t}(C_{3})_{036}(\mathbf{\mathbf{k},\mathbf{k}^{\prime \prime \prime },%
\mathbf{k}^{\prime }+k}^{\prime \prime },t^{\prime }))  \notag \\
& -\kappa _{51}(\mathbf{\mathbf{k}^{\prime }})(\kappa _{06}(\mathbf{\mathbf{k%
}})(C_{4})_{6235}(\mathbf{\mathbf{k},k}^{\prime \prime },\mathbf{\mathbf{k}%
^{\prime \prime \prime },\mathbf{k}^{\prime },}t^{\prime })+\partial
_{t}(C_{4})_{0235}(\mathbf{\mathbf{k},k}^{\prime \prime },\mathbf{\mathbf{k}%
^{\prime \prime \prime },\mathbf{k}^{\prime },}t^{\prime }))  \notag \\
& -\frac{1}{3}(\kappa _{05}(\mathbf{\mathbf{k}})\partial _{t}(C_{4})_{5123}(%
\mathbf{\mathbf{k},\mathbf{k}^{\prime },k}^{\prime \prime },\mathbf{\mathbf{k%
}^{\prime \prime \prime },}t^{\prime })+\partial _{tt}(C_{4})_{0123}((%
\mathbf{\mathbf{k},\mathbf{k}^{\prime },k}^{\prime \prime },\mathbf{\mathbf{k%
}^{\prime \prime \prime },}t^{\prime }))  \label{a.47}
\end{align}%
\begin{align}
& \Phi 2_{0123}(\mathbf{\mathbf{k}^{\prime },\mathbf{k}^{\prime \prime },k}%
^{\prime \prime \prime },t^{\prime },t^{\prime \prime })=  \notag \\
& =2\gamma _{51}(\mathbf{\mathbf{k}^{\prime }})(\kappa _{06}(\mathbf{\mathbf{%
k}})(C_{3})_{637}(\mathbf{\mathbf{k},\mathbf{k}^{\prime \prime \prime },%
\mathbf{k}^{\prime }+k}^{\prime \prime },t^{\prime })+\partial
_{t}(C_{3})_{037}(\mathbf{\mathbf{k},\mathbf{k}^{\prime \prime \prime },%
\mathbf{k}^{\prime }+k}^{\prime \prime },t^{\prime }))\times  \notag \\
& \times (\kappa _{78}(\mathbf{\mathbf{k}^{\prime }+k}^{\prime \prime
})(C_{3})_{825}(\mathbf{\mathbf{k}^{\prime }+k}^{\prime \prime },\mathbf{k}%
^{\prime \prime },\mathbf{\mathbf{k}^{\prime },}t^{\prime \prime })+\partial
_{t}(C_{3})_{725}((\mathbf{\mathbf{k}^{\prime }+k}^{\prime \prime },\mathbf{k%
}^{\prime \prime },\mathbf{\mathbf{k}^{\prime },}t^{\prime \prime }))  \notag
\\
& +2(\kappa _{05}(\mathbf{\mathbf{k}})(C_{3})_{524}(\mathbf{\mathbf{k},%
\mathbf{k}^{\prime \prime },\mathbf{k}^{\prime }+k}^{\prime \prime \prime
},t^{\prime })+\partial _{t}(C_{3})_{024}(\mathbf{\mathbf{k},\mathbf{k}%
^{\prime \prime },\mathbf{k}^{\prime }+k}^{\prime \prime \prime },t^{\prime
}))\times  \notag \\
& \times (\omega _{16}(\mathbf{\mathbf{k}^{\prime }})(\kappa _{47}(\mathbf{%
\mathbf{k}^{\prime }+k}^{\prime \prime \prime })(C_{3})_{763}(\mathbf{%
\mathbf{k}^{\prime }+k}^{\prime \prime \prime },\mathbf{k}^{\prime },\mathbf{%
\mathbf{k}^{\prime \prime \prime },}t^{\prime \prime })+\partial
_{t}(C_{3})_{463}(\mathbf{\mathbf{k}^{\prime }+k}^{\prime \prime \prime },%
\mathbf{k}^{\prime },\mathbf{\mathbf{k}^{\prime \prime \prime },}t^{\prime
\prime }))  \notag \\
& +\kappa _{47}(\mathbf{\mathbf{k}^{\prime }+k}^{\prime \prime \prime
})\partial _{t}(C_{3})_{713}(\mathbf{\mathbf{k}^{\prime }+k}^{\prime \prime
\prime },\mathbf{k}^{\prime },\mathbf{\mathbf{k}^{\prime \prime \prime },}%
t^{\prime \prime })+\frac{1}{2}\partial _{tt}(C_{3})_{413}(\mathbf{\mathbf{k}%
^{\prime }+k}^{\prime \prime \prime },\mathbf{k}^{\prime },\mathbf{\mathbf{k}%
^{\prime \prime \prime },}t^{\prime \prime }))  \label{a.48}
\end{align}%
\begin{align}
& (K_{2})_{01}\left( \mathbf{\mathbf{k}^{\prime },\mathbf{k}}^{\prime \prime
},t^{\prime }\right) =\gamma _{01}(\mathbf{\mathbf{k}})(2\pi )^{3}\delta (%
\mathbf{k}^{\prime \prime })\langle h_{2}\rangle \delta (t^{\prime })  \notag
\\
& +\gamma _{31}(\mathbf{\mathbf{k}^{\prime }})(\partial _{t^{\prime
}}(C_{3})_{032}(\mathbf{\mathbf{k},\mathbf{k}^{\prime },k}^{\prime \prime
},t^{\prime })+\kappa _{04}(\mathbf{\mathbf{k}})(C_{3})_{432}(\mathbf{%
\mathbf{k},\mathbf{k}^{\prime },k}^{\prime \prime },t^{\prime }))  \notag \\
& +\partial _{t^{\prime }t^{\prime }}(C_{3})_{012}(\mathbf{\mathbf{k},%
\mathbf{k}^{\prime },k}^{\prime \prime },t^{\prime })+i\,k_{c}^{\prime
\prime }\partial _{t^{\prime }}(C_{3})_{01c}(\mathbf{\mathbf{k},\mathbf{k}%
^{\prime },k}^{\prime \prime },t^{\prime })+\omega _{14}(\mathbf{\mathbf{k}%
^{\prime }})\partial _{t^{\prime }}(C_{3})_{042}(\mathbf{\mathbf{k},\mathbf{k%
}^{\prime },k}^{\prime \prime },t^{\prime })  \notag \\
& +\kappa _{03}(\mathbf{\mathbf{k}})(\partial _{t^{\prime }}(C_{3})_{312}(%
\mathbf{\mathbf{k},\mathbf{k}^{\prime },k}^{\prime \prime },t^{\prime
})+i\,k_{c}^{\prime \prime }(C_{3})_{31c}(\mathbf{\mathbf{k},\mathbf{k}%
^{\prime },k}^{\prime \prime },t^{\prime })+\omega _{14}(\mathbf{\mathbf{k}%
^{\prime }})(C_{3})_{342}(\mathbf{\mathbf{k},\mathbf{k}^{\prime },k}^{\prime
\prime },t^{\prime }))  \label{a.49}
\end{align}

The wave number $\mathbf{k}$ is, in (\ref{a.47}) and (\ref{a.48}), to be
substituted by $\mathbf{\mathbf{k}^{\prime }+\mathbf{k}^{\prime \prime }+k}%
^{\prime \prime \prime }$,\ and in (\ref{a.49}) by $\mathbf{\mathbf{k}%
^{\prime }+\mathbf{k}^{\prime \prime }}$. - The calculation in sec. 8 of the
kernel function $\Delta _{3}$\ appearing in (\ref{7.1}) yields the
expression (again $\mathbf{k=\mathbf{k}^{\prime }+\mathbf{k}^{\prime \prime
}+k}^{\prime \prime \prime }$ ):

\begin{align}
\Delta _{3}& (\mathbf{k},\mathbf{k}^{\prime },\mathbf{k}^{\prime \prime },%
\mathbf{k}^{\prime \prime \prime })=-\tfrac{1}{3}(s_{4})_{0123}(\mathbf{k}) 
\notag \\
& +\gamma _{04}(\mathbf{k})(\tfrac{1}{3}(j_{4})_{4123}-(2\pi )^{3}\delta (%
\mathbf{k}-\mathbf{k}^{\prime })\delta _{14}\delta _{23})  \notag \\
& +2(s_{3})_{056}(\mathbf{k})(s_{134}(\mathbf{k}^{\prime })-(s_{3})_{413}(%
\mathbf{k}^{\prime }+\mathbf{k}^{\prime \prime \prime }))(\zeta
_{2})_{5624}^{-1}(\mathbf{k}^{\prime \prime },\mathbf{k}^{\prime }+\mathbf{k}%
^{\prime \prime \prime })  \notag \\
& -2(s_{3})_{056}(\mathbf{k})(j_{734}\gamma _{41}(\mathbf{k}^{\prime })-%
\tfrac{1}{4}(j_{3})_{413}\gamma _{47}(\mathbf{k}^{\prime }+\mathbf{k}%
^{\prime \prime \prime }))(\zeta _{2})_{5627}^{-1}(\mathbf{k}^{\prime \prime
},\mathbf{k}^{\prime }+\mathbf{k}^{\prime \prime \prime })  \notag \\
& +2(s_{3})_{047}(\mathbf{k})(s_{3})_{856}(\mathbf{k}^{\prime }+\mathbf{k}%
^{\prime \prime })(\zeta _{2})_{4738}^{-1}(\mathbf{k}^{\prime \prime \prime
},\mathbf{k}^{\prime }+\mathbf{k}^{\prime \prime })(\zeta _{2})_{5629}^{-1}(%
\mathbf{k}^{\prime \prime },\mathbf{k}^{\prime })\kappa _{19}(\mathbf{k}%
^{\prime })  \notag \\
& +\frac{1}{2}(s_{3})_{047}(\mathbf{k})(s_{3})_{589}(\mathbf{k}^{\prime }+%
\mathbf{k}^{\prime \prime })(\zeta _{2})_{4736}^{-1}(\mathbf{k}^{\prime
\prime \prime },\mathbf{k}^{\prime }+\mathbf{k}^{\prime \prime })(\zeta
_{2})_{8921}^{-1}(\mathbf{k}^{\prime \prime },\mathbf{k}^{\prime })\kappa
_{56}(\mathbf{k}^{\prime }+\mathbf{k}^{\prime \prime })  \notag \\
& +\left( -\frac{1}{(2\pi )^{3}}\int d\mathbf{\breve{k}\,}\Phi _{47856}(%
\mathbf{\breve{k}},-\mathbf{\breve{k}})(s_{3})_{056}(\mathbf{\breve{k}},-%
\mathbf{\breve{k}})+(s_{4})_{0478}(\mathbf{k})\right) \times  \notag \\
& \text{ \ \ \ \ \ \ \ \ \ \ \ \ \ \ \ \ \ \ \ \ \ \ \ \ \ \ \ \ \ \ \ \ \ \
\ \ \ \ \ \ \ \ \ \ \ \ \ \ \ \ \ \ \ \ \ \ \ \ \ \ \ \ \ \ \ \ \ \ \ \ }%
\times (\zeta _{3})_{478239}^{-1}(\mathbf{k}^{\prime \prime },\mathbf{k}%
^{\prime \prime \prime },\mathbf{k}^{\prime })\kappa _{19}(\mathbf{k}%
^{\prime })  \notag \\
& -\tfrac{1}{2}(s_{3})_{047}(\mathbf{k})(s_{3})_{568}(\mathbf{k}^{\prime }+%
\mathbf{k}^{\prime \prime \prime })(\zeta _{2})_{472\bar{9}}^{-1}(\mathbf{k}%
^{\prime \prime },\mathbf{k}^{\prime }+\mathbf{k}^{\prime \prime \prime
})\times  \notag \\
& \text{ \ \ \ \ \ \ \ \ \ \ \ \ \ \ \ \ \ \ \ \ \ \ \ \ \ \ \ \ \ \ \ \ }%
\times (\check{\zeta}_{3})_{913568}^{-1}(\mathbf{k}^{\prime }+\mathbf{k}%
^{\prime \prime \prime },\mathbf{k}^{\prime },\mathbf{k}^{\prime \prime
\prime })\kappa _{\bar{9}9}(\mathbf{k}^{\prime }+\mathbf{k}^{\prime \prime
\prime })  \notag \\
& +(s_{3})_{047}(\mathbf{k})\left( \kappa _{19}(\mathbf{k}^{\prime })\frac{1%
}{(2\pi )^{3}}\int d\mathbf{\breve{k}\,}\Phi _{23956}(\mathbf{\breve{k}},-%
\mathbf{\breve{k}})(\zeta _{2})_{4756}^{-1}(\mathbf{\breve{k}},-\mathbf{%
\breve{k}})+\right.  \notag \\
& \text{ \ \ \ \ \ \ \ \ \ \ \ \ \ \ \ \ \ \ \ \ \ \ \ \ \ \ \ \ \ \ }\left.
+\tfrac{1}{2}(j_{3})_{813}(\zeta _{2})_{4729}^{-1}(\mathbf{k}^{\prime \prime
},\mathbf{k}^{\prime }+\mathbf{k}^{\prime \prime \prime })\kappa _{98}(%
\mathbf{k}^{\prime }+\mathbf{k}^{\prime \prime \prime })\right)  \notag \\
& +\tfrac{1}{2}(s_{3})_{047}(\mathbf{k})(s_{3})_{\bar{5}\bar{6}\bar{7}}(%
\mathbf{k}^{\prime }+\mathbf{k}^{\prime \prime \prime })(\zeta
_{2})_{1389}^{-1}(\mathbf{k}^{\prime },\mathbf{k}^{\prime \prime \prime
})(\zeta _{2})_{472\bar{8}}^{-1}(\mathbf{k}^{\prime \prime },\mathbf{k}%
^{\prime }+\mathbf{k}^{\prime \prime \prime })\times  \notag \\
& \text{ \ \ \ \ \ \ \ \ \ \ \ \ \ \ \ \ }\times (\check{\zeta}_{3})_{\bar{4}%
89\bar{5}\bar{6}\bar{7}}^{-1}(\mathbf{k}^{\prime }+\mathbf{k}^{\prime \prime
\prime },\mathbf{k}^{\prime },\mathbf{k}^{\prime \prime \prime })\kappa _{%
\bar{8}\bar{9}}(\mathbf{k}^{\prime }+\mathbf{k}^{\prime \prime \prime
})\kappa _{\bar{9}\bar{4}}(\mathbf{k}^{\prime }+\mathbf{k}^{\prime \prime
\prime })  \label{a.50}
\end{align}

With, in sec. 9, several of the static correlations found to be zero, the
formula simplifies:

\begin{align}
\Delta _{3}& (\mathbf{k},\mathbf{k}^{\prime },\mathbf{k}^{\prime \prime },%
\mathbf{k}^{\prime \prime \prime })=\gamma _{04}(\mathbf{k})(\tfrac{1}{3}%
(j_{4})_{4123}-(2\pi )^{3}\delta (\mathbf{k}-\mathbf{k}^{\prime })\delta
_{14}\delta _{23})  \notag \\
& +2(s_{3})_{056}(\mathbf{k})((s_{3})_{134}(\mathbf{k}^{\prime
})-(s_{3})_{413}(\mathbf{k}^{\prime }+\mathbf{k}^{\prime \prime \prime
}))(\zeta _{2})_{5624}^{-1}(\mathbf{k}^{\prime \prime },\mathbf{k}^{\prime }+%
\mathbf{k}^{\prime \prime \prime })  \notag \\
& +2(s_{3})_{047}(\mathbf{k})(s_{3})_{856}(\mathbf{k}^{\prime }+\mathbf{k}%
^{\prime \prime })(\zeta _{2})_{4738}^{-1}(\mathbf{k}^{\prime \prime \prime
},\mathbf{k}^{\prime }+\mathbf{k}^{\prime \prime })(\zeta _{2})_{5629}^{-1}(%
\mathbf{k}^{\prime \prime },\mathbf{k}^{\prime })\kappa _{19}(\mathbf{k}%
^{\prime })  \notag \\
& +\frac{1}{2}(s_{3})_{047}(\mathbf{k})(s_{3})_{589}(\mathbf{k}^{\prime }+%
\mathbf{k}^{\prime \prime })(\zeta _{2})_{4736}^{-1}(\mathbf{k}^{\prime
\prime \prime },\mathbf{k}^{\prime }+\mathbf{k}^{\prime \prime })(\zeta
_{2})_{8921}^{-1}(\mathbf{k}^{\prime \prime },\mathbf{k}^{\prime })\kappa
_{56}(\mathbf{k}^{\prime }+\mathbf{k}^{\prime \prime })  \notag \\
& -\tfrac{1}{2}(s_{3})_{047}(\mathbf{k})(s_{3})_{568}(\mathbf{k}^{\prime }+%
\mathbf{k}^{\prime \prime \prime })(\zeta _{2})_{472\bar{9}}^{-1}(\mathbf{k}%
^{\prime \prime },\mathbf{k}^{\prime }+\mathbf{k}^{\prime \prime \prime
})\times  \notag \\
& \text{ \ \ \ \ \ \ \ \ \ \ \ \ \ \ \ \ \ \ \ \ \ \ \ \ \ \ \ \ \ \ \ \ }%
\times (\check{\zeta}_{3})_{913568}^{-1}(\mathbf{k}^{\prime }+\mathbf{k}%
^{\prime \prime \prime },\mathbf{k}^{\prime },\mathbf{k}^{\prime \prime
\prime })\kappa _{\bar{9}9}(\mathbf{k}^{\prime }+\mathbf{k}^{\prime \prime
\prime })  \notag \\
& +\tfrac{1}{2}(s_{3})_{047}(\mathbf{k})(s_{3})_{\bar{5}\bar{6}\bar{7}}(%
\mathbf{k}^{\prime }+\mathbf{k}^{\prime \prime \prime })(\zeta
_{2})_{1389}^{-1}(\mathbf{k}^{\prime },\mathbf{k}^{\prime \prime \prime
})(\zeta _{2})_{472\bar{8}}^{-1}(\mathbf{k}^{\prime \prime },\mathbf{k}%
^{\prime }+\mathbf{k}^{\prime \prime \prime })\times  \notag \\
& \text{ \ \ \ \ \ \ \ \ \ \ \ \ \ \ \ \ }\times (\check{\zeta}_{3})_{\bar{4}%
89\bar{5}\bar{6}\bar{7}}^{-1}(\mathbf{k}^{\prime }+\mathbf{k}^{\prime \prime
\prime },\mathbf{k}^{\prime },\mathbf{k}^{\prime \prime \prime })\kappa _{%
\bar{8}\bar{9}}(\mathbf{k}^{\prime }+\mathbf{k}^{\prime \prime \prime
})\kappa _{\bar{9}\bar{4}}(\mathbf{k}^{\prime }+\mathbf{k}^{\prime \prime
\prime })  \label{a.51}
\end{align}


\begin{thebibliography}{99}
\bibitem{Ho} E. Hopf, J. ratl. mech. Anal. 1, p. 87, 1952

\bibitem{McCo} W. D. McComb, The physics of fluid turbulence, Oxford Science
Pub., Oxford 1990

\bibitem{DNS} P. Moin, K. Mahesh: Direct numerical simulation. A tool in
turbulence research. In: Annual Review of Fluid Mechanics. Vol. 30, 1998, S.
539--578

\bibitem{Hu} K. Huang, Statistical Mechanics. Wiley, N. Y. 1963

\bibitem{gra} H. Grabert, Projection operator techniques in nonequilibrium
statistical mechanics. Springer, Berlin, Heidelberg, New York (1982)

\bibitem{pi03} J. Piest, arXiv physics/0310054, 2003

\bibitem{Ka} K. Kawasaki, Ann. Phys. 61, p. 1, 1970;

P. C. Martin, E. D. Siggia, H. A. Rose, Phys. Rev. A8, p. 423, 1973;

U. Deker, F. Haake, Phys. Rev. A11, 2043, 1975

\bibitem{scho} J. Schofield, R. Lim, I. Oppenheim, Physica A181, p. 89, 1992;

\bibitem{vzs} R. van Zon, J. Schofield, Phys. Rev. E65, 011106, 2001

\bibitem{pi07} J. Piest, arXiv 0711.2790v1, 2007;

- , arXiv 0803.3972, 2008

\bibitem{zu} Zubarev, D.; Mozorov, V.; R\"{o}pke, G.: Statistical mechanics
of nonequilibrium processes. Vol. 1, Akademie Verlag Berlin 1996.

\bibitem{k} K. Kawasaki, Ann. Phys. 61, (1970), 1

\bibitem{ehl} M. H. Ernst, E. H. Hauge, J. M. J. van Leeuwen, Phys. Rev. A
4, 5, (1971), 2055

\bibitem{mu} A. M\"{u}nster: Statistische Thermodynamik. Springer, Berlin
1956

\bibitem{sch} H. Schlichting, Boundary layer theory, 7$^{th}$ ed., McGraw
Hill 1979

\bibitem{z85} E. Zauner: Visualization of the viscous flow induced by a
round jet. J. Fluid Mech. 154, 111-119, 1985

\bibitem{sch85} W. Schneider: Decay of momentum flux in submerged jets. J.
Fluid Mech. 154, 91-110, 1985

\bibitem{szb87} W. Schneider, E. Zauner, H. B\"{o}hm:\ The recirculatory
flow induced by a laminar axixymmetric jet issuing from al wall. J. Fluids
Eng. 109, 237-241, 1987
\end{thebibliography}
\end{document}